\documentclass[preprint]{revtex4}

\usepackage{amsmath, amssymb, amsthm}
\usepackage{amsfonts}
\usepackage{graphicx}
\usepackage{palatino}
\usepackage{url}

\begin{document}

\begin{center}
\textsc{The dynamic of information-driven coordination phenomena: a transfer entropy analysis}
\end{center}

\noindent{Javier Borge-Holthoefer,$^{1\ast}$ Nicola Perra,$^{2\ast}$ Bruno Gon\c{c}alves,$^{3}$ Sandra Gonz\'alez-Bail\'on,$^{4}$ Alex Arenas,$^{5}$ Yamir Moreno,$^{6,7,8\ast}$ Alessandro Vespignani$^{2,8,9\ast}$}

\noindent{\normalsize{$^{1}$Qatar Computing Research Institute, HBKU, Doha, Qatar}}\\
\normalsize{$^{2}$Laboratory for the Modeling of Biological and Socio-technical Systems, Northeastern University, Boston 02115, USA\\
\normalsize{$^{3}$Aix Marseille Universit\'e, Universit\'e de Toulon, CNRS, CPT, UMR 7332, 13288 Marseille, France\\
\normalsize{$^{4}$Annenberg School for Communication, University of Pennsylvania, Philadelphia 19104, USA\\
\normalsize{$^{5}$Departament d'Enginyeria Inform\`{a}tica i Matem\`{a}tiques, Universitat Rovira i Virgili, 43007 Tarragona, Spain\\
\normalsize{$^{6}$Institute for Biocomputation and Physics of Complex Systems (BIFI), University of Zaragoza,}\\
\normalsize{50018 Zaragoza, Spain}\\
\normalsize{$^{7}$Department of Theoretical Physics, University of Zaragoza, Zaragoza 50009, Spain}\\
\normalsize{$^{8}$ISI Foundation, Turin, Italy}\\
\normalsize{$^{9}$Institute for Quantitative Social Sciences at Harvard University, Cambridge MA 02138, USA\\
\\
\normalsize{$^\ast$To whom correspondence should be addressed: jborge@qf.org.qa, n.perra@neu.edu, yamir.moreno@gmail.com, a.vespignani@neu.edu}


\begin{abstract}
Data from social media are providing unprecedented opportunities to investigate the processes that rule the dynamics of collective social phenomena.  Here, we consider an information theoretical approach to define and measure the temporal and structural signatures typical of collective social events as they arise and gain prominence.  We use the symbolic transfer entropy analysis of micro-blogging time series to extract directed networks of influence among geolocalized sub-units in social systems. This methodology captures the emergence of system-level dynamics close to the onset of socially relevant collective phenomena. The framework is validated against a detailed empirical analysis of five case studies. In particular, we identify a change in the characteristic time-scale of the information transfer that flags the onset of information-driven collective phenomena. Furthermore, our approach identifies an order-disorder transition in the directed network of influence between social sub-units. In the absence of a clear exogenous driving, social collective phenomena can be represented as endogenously-driven structural transitions of the information transfer network. This study provides results that can help define models and predictive algorithms for the analysis of societal events based on open source data. 
\end{abstract}

\maketitle

A vivid scientific and popular media debate has recently centered on the role that social networking tools play in coordinating collective phenomena. Examples include street protests, civil unrests, consensus formation, or the emergence of electoral preferences. A flurry of studies have analyzed the correlation of search engine queries, microblogging posts and other open data sources with the incidence of infectious disease \cite{culotta2010towards,ginsberg2009detecting,10.1371/journal.pcbi.1004239,chakraborty2014forecasting}, box office returns~\cite{Asur2010}, stock market behavior \cite{bollen2011twitter,Curme2014}, election outcomes \cite{tumasjan2010predicting,livne2011party}, popular votes results~\cite{ciulla2012beating}, crowd sizes \cite{botta2015quantifying}, and social unrest~\cite{xu:2014,Ramakrishnan:2014}. Many other studies, however, have also pointed out the challenges big data presents and the likely methodological pitfalls that might result from their analysis \cite{skoric2012tweets,sang2012predicting,ratkiewicz2011detecting,gayo2012wanted,tufekci2014big,lazer2014parable,helbing2013globally}. 
This prior work suggests that more research is needed to develop methods for exploiting the value of social media data while overcoming their limitations. 

Here, we use micro-blogging data to extract networks of causal influence among different geographical sub-units before, during, and after collective social phenomena. In order to ground our work on empirical data, we analyze five datasets that track Twitter communications around five well-known social events: the release of a Hollywood blockbuster movie; two massive political protests; the discovery of the Higgs boson; and the acquisition of Motorola by Google. We selected these case studies because they represent different points in a theoretical continuum that separates two types of collective phenomena: those that can be represented as an endogenously-driven exchange of information; and those that respond more clearly to factors that are exogenous to the system. In our context, these phenomena refer to dynamics of information exchange through social media: in some cases, discussions evolve organically, building up momentum up to the point where the exchange of information is generalized; in some other cases, however, the discussions emerge suddenly as a reaction to some unexpected external event~\cite{lehmann2012dynamical}. 

For each case study we adopt the transfer entropy approach to define an effective social connectivity at the macro-scale, and study the coordinated activation of localized populations. We address two foundational problems: first, the identification of the characteristic time-scale of social events as they develop, gather force, and burst into generalized attention; and second, the representation of the structural signature typical of the communication dynamics that underlie social phenomena. We find that the onset of social collective phenomena are characterized by a drop of the characteristic time-scale; we also show that the emergence of coherent patterns of information flow can be mapped into order-disorder transitions in the underlying connectivity patterns of the transfer entropy network. The methodology we present here can therefore be used to gain new insights on the structural and functional relations occurring in large-scale structured populations, eventually leading to the identification of metrics that might be used for the definition of precursors of large-scale social events.

\section{Results}
We consider the dataset concerning the time stamped and geolocalized time-series of tweets associated to the following events: the Spanish 15M social unrest in 2011;  the {\em Outono Brasileiro} (``Brazilian Autumn'') in 2013;  the discovery of the Higgs boson in 2012;  the release of an Hollywood blockbuster in 2012;  the acquisition of Motorola by Google in 2011.
All datasets cover a time-span preceding and following the event and details on data collection, including keyword selection and the geolocalization of messages, can be found in the Materials and Methods section and in the Supplementary Information (SI).

\begin{figure*}[h]
\begin{center}
\includegraphics[width=0.95\textwidth]{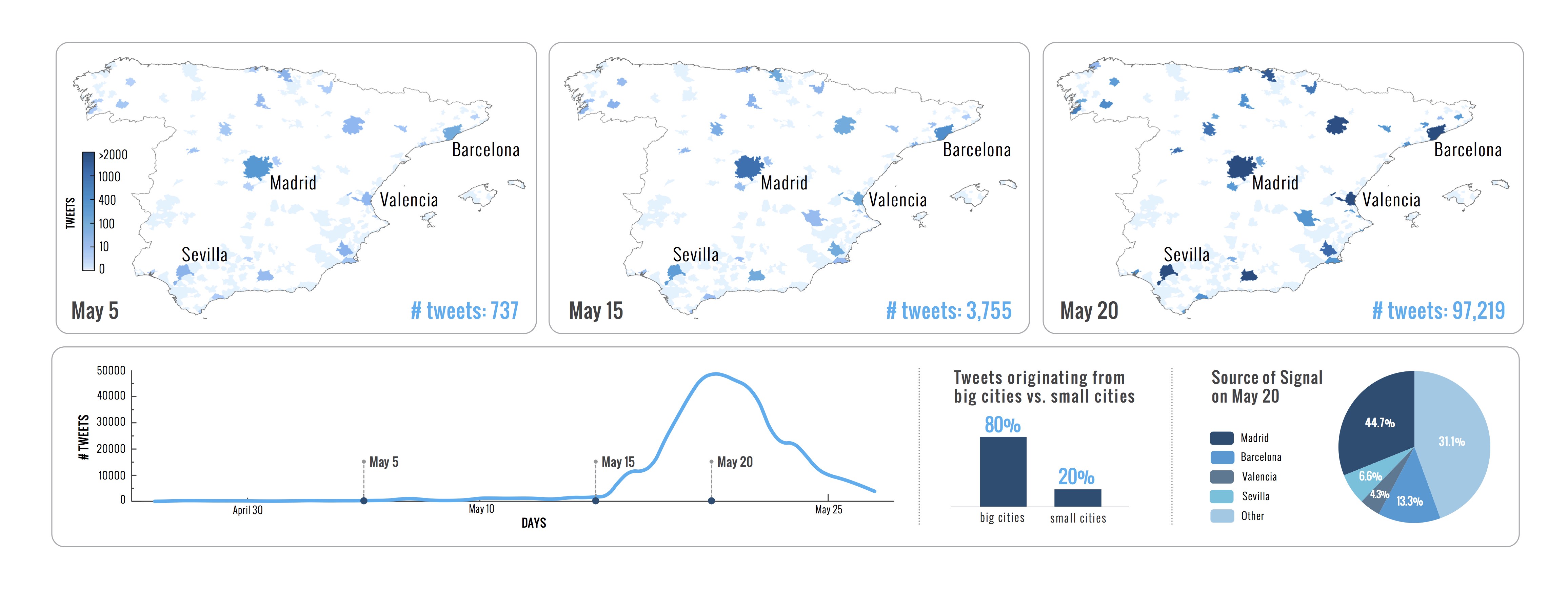}
\end{center}
\caption{\label{fig1}{\bf Spatio-temporal activity as observed from the microblogging platform Twitter.} Spain's 15M protest growth in time shows that the protest did not transcend the online sphere until May $15^{th}$ when the political movement emerged on the streets. Broadcasting traditional media started reporting on it soon after; by that time, demonstrations had been held in the most important cities of the country.}
\end{figure*}

The spatio-temporal annotation of each tweet in the time series allows the construction of spatially localized activity maps that help identify, as time unfolds, the role that different geographical sub-units played in the global exchange of information. For each dataset the definition of the corresponding spatial unit is performed according to administrative and geographical boundaries as specified in the Material and Methods section. The time-stamped series of tweets originated from each spatial sub-unit $X$ (supra-urban aggregates) defines the activity time series $X_t$  of the corresponding sub-unit in the social system. Timestamps are modified for each dataset to account for different time zones (see SI for details). 

Activity time series encode the role of each geographical sub-unit, a sort of {\em who-steers-whom}, and several techniques can be used to detect directed exchange of information across the social system. Here, we characterize the dominating direction of information flow between spatial sub-units using Symbolic Transfer Entropy (STE) \cite{staniek08,bandt2002permutation}. This well-established technique has been used to infer directional influence between dynamical systems \cite{schreiber2000measuring,hlavavckova2007causality,ni2014information} and to analyze patterns of brain connectivity \cite{lizier2011multivariate}. 

Symbolic transfer entropy quantifies the directional flow of information between two time series $X$ and $Y$ by, first, categorizing the signals in a small set of symbols or alphabet (see Figure S2 of the SI); and, then, computing from the relative frequency of symbols in each sequence $\hat{X}$ and $\hat{Y}$ the joint and conditional probabilities of the sequences indices as 
\begin{equation}
T_{Y,X}=\sum p\left(\hat{x}_{i+\delta},\hat{x}_{i},\hat{y}_{i}\right) \log_2\left(\frac{p\left(\hat{x}_{i+\delta}|\hat{x}_{i},\hat{y}_{i}\right)}{p\left(\hat{x}_{i+\delta}|\hat{x}_{i}\right)}\right)
\end{equation}
where the sum runs over all symbols and $\delta=1$. The transfer entropy refers to the deviations of the cross-Markovian property of the series (independence between them), measured as the Kullback-Leibler divergence \cite{kullback1951information} (see the SI for all technical details). An important feature of symbolic approaches is that it discounts the relative magnitude of each time series; this is important in our case because different geographical units differ largely in population density or internet penetration rates.

Within this framework, we first analyze the temporal patterns characterizing the flow of information. Admittedly, micro-blogging data can be sampled at different time-scales $\Delta t$. In order to select the optimal sampling rate we consider all possible pairs $\left(X, Y\right)$ of geographical units and measure the total STE in the system $T=\sum_{XY}T_{X,Y}$  as a function of $\Delta t$. We consider the system-wide characteristic sampling time-scale $\tau$ as that which maximizes the total information flow $T$. This quantity provides an indication of the time-scale at which the information is being exchanged in the system, not necessarily correlated with volume. Interestingly, the characteristic time-scale $\tau$ changes as the phenomena under analysis unfold, i.e. it decreases as the system approaches the exponential increase in overall activity that signals the onset of the collective phenomena. As shown in the top panels of Figure~\ref{fig2}, $\tau$ is a proxy for the internally generated coordination in the system that culminates at the very same time of the occurrence of the social event: the street protest day, in the case of political unrest; the movie release date, in the case of the Hollywood blockbuster; and the announcement to the press of the Higgs boson discovery. The only clear exception to this behavior is offered by the company acquisition dataset: the Google-Motorola announcement is a clear example of collective phenomena that is driven mostly by an exogenous factor, i.e. a media announcement. In this case, the dynamical time-scale is constant until the announcement is made public. In the SI we present the same analysis for the randomized signals, showing that time-scale variations are, as expected, washed out from the signal.

The maximized information exchange can be analyzed at the level of geographical subunits by constructing the effective directed network \cite{sporns2004organization} of information flow on a daily basis. This network is encoded in the matrix $\left\{T_{XY}\right\}$ that contains pairwise information about how each component in the system controls (or is controlled by) the others. The matrix $\left\{T_{XY}\right\}$ is asymmetric. The directionality is crucial and captures that the geographic area $x$ can exert some driving on area $y$, and at the same time $y$ might exert some driving on $x$. For this reason it is convenient to define the directionality index $T_{X,Y}^S=T_{Y,X}-T_{X,Y}$ measuring the balance of information flow in both directions. This index quantifies the dominant direction of information flow and is expected to have positive values for unidirectional couplings with $x$ as the driver and negative values if $y$ is driving $x$. For symmetric bidirectional couplings we expect $T_{X,Y}^S$ to be null.

\begin{figure}
\begin{center}
\includegraphics[width=.9\columnwidth]{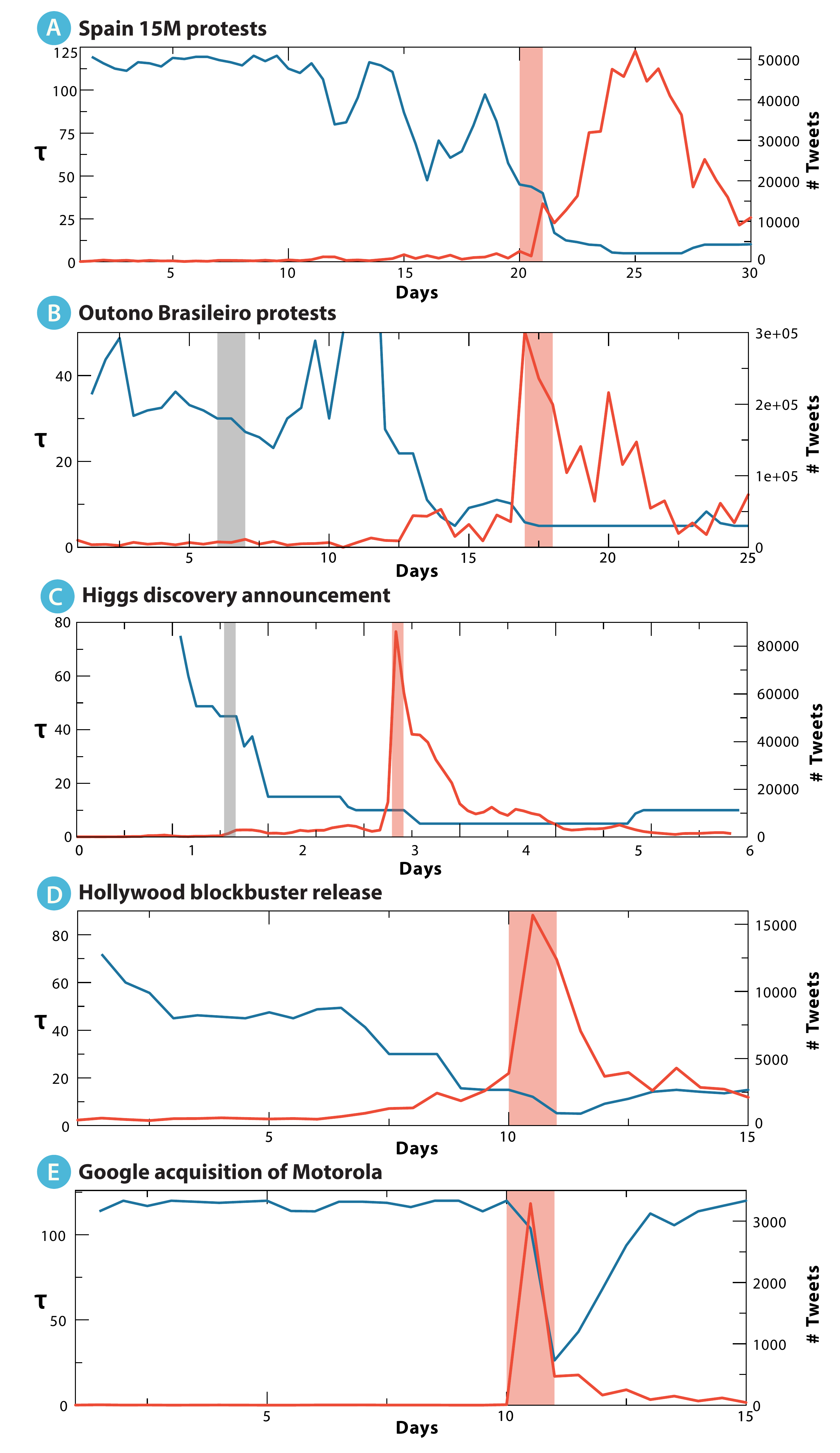}
\end{center}
\caption{\label{fig2}{\bf Characteristic time-scale $\tau$.} The panels report the variation of the characteristic time-scale (blue) that maximizes the $STE$ flow as the social event is approached. Red lines correspond to activity volume (number of tweets). Light red vertical lines correspond to the onset of the main social event, gray ones (in B and C) indicate a smaller precursor event: A) the 15M event shows a progressive decline of the characteristic time-scale well before the actual social event; the same is observed for the Outono Brazileiro in B) (note a data blackout between days 10 and 11). The patterns for the Higgs boson discovery dataset in C) and the Hollywood blockbuster data D) reveal also a drop in the characteristic time-scale, although this is smoother in the movie case. Finally, E) the Google-Motorola deal triggers a high volume of microblogging activity without actual change in the time-scale of the information flow. In this case the decline is observed in the aftermath of the announcement. As discussed in the text, this event is the only one that is clearly elicited by an exogenous trigger.}
\end{figure}

\begin{figure*}
\begin{center}
\includegraphics[width=0.9\textwidth]{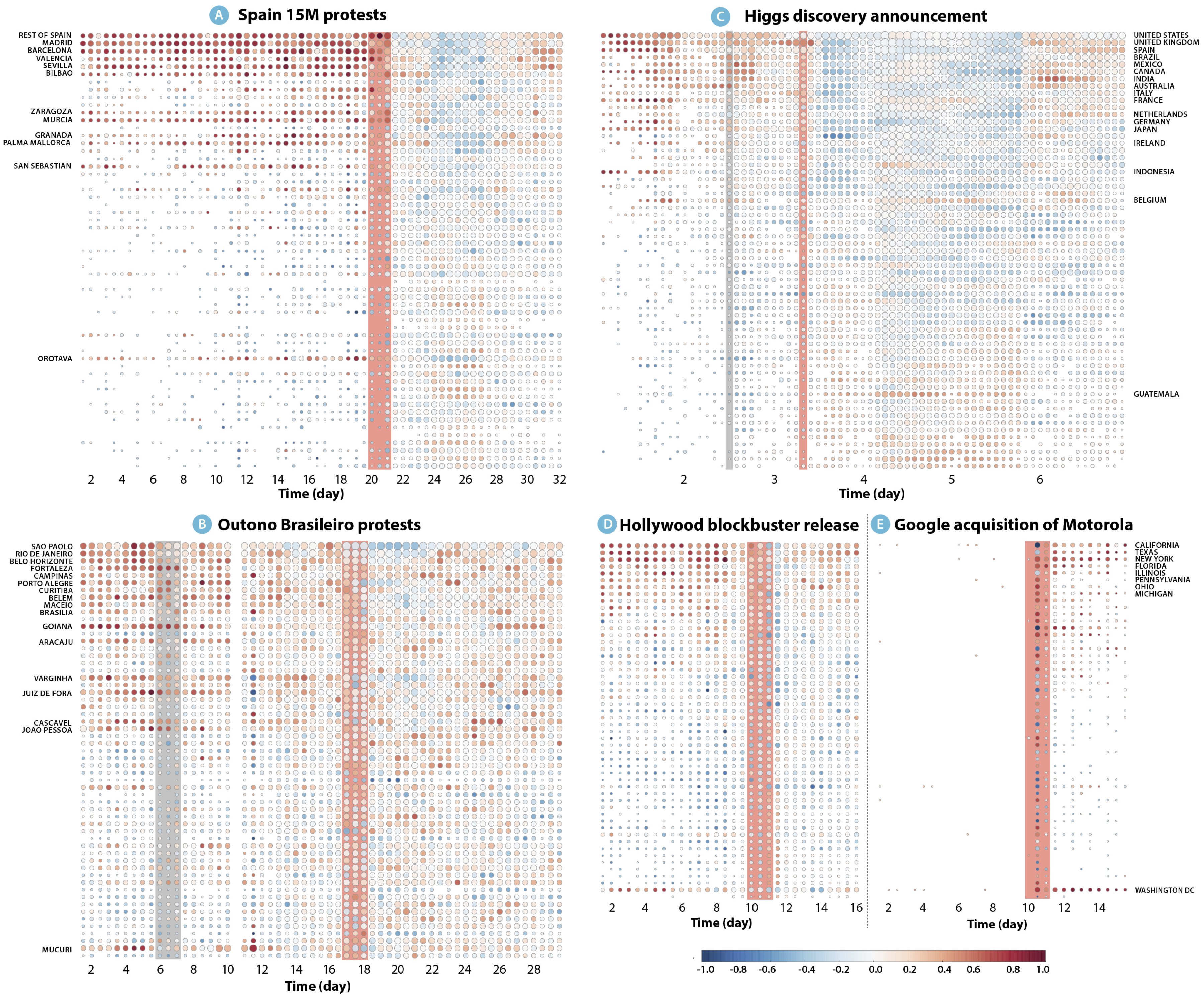}
\end{center}
\caption{\label{fig3}{\bf Evolution of information flow balance between geographical locations for the analyzed events.} The color goes from dark blue to dark red (white corresponds to null driving), with the former standing for negative values of $\sum_yT_{X,Y}^S$  (e.g., driven locations) and the latter corresponding to positive information flow balances (i.e., drivers). The size of the circles is log-proportional to the number of messages sent from the location at that time and the vertical bars mark the day of the main event. The geographical locations are ordered according to population size, except for C, in which countries are ranked with the amount of Higgs-related tweets produced.}
\end{figure*}

Figure~\ref{fig3} reports the temporal evolution of the maximized $\sum_YT_{X,Y}^S$  that provides the information flow balance of each specific geographical area. The results show that in the 15M grassroots protests, a limited number of urban areas are initially driving the onset of the social phenomena. These units can mostly be identified with major cities; however the analysis also uncovers {\em hidden} drivers, such as Orotava, a less known urban area. Only after the first demonstration day on May $15^{th}$ the driving role becomes much more homogeneously distributed. In the Brazilian case, a set of clear drivers is present only during the onset phase preceding a demonstration on June $6^{th}$, becoming fuzzier up to the major demonstration (June $17^{th}$) and totally blurred afterwards. We find a similar behavior in the Higgs boson cases (with rumors around the discovery on July $2^{nd}$ and final announcement on July $4^{th}$) \cite{de2013anatomy}. The blockbuster case is driven by a steady excitement of the public before the movie release. Again, as expected, we observe completely different patterns in the case of the Google dataset. 

\begin{figure*}
\centering
\includegraphics[width=\columnwidth,clip=0]{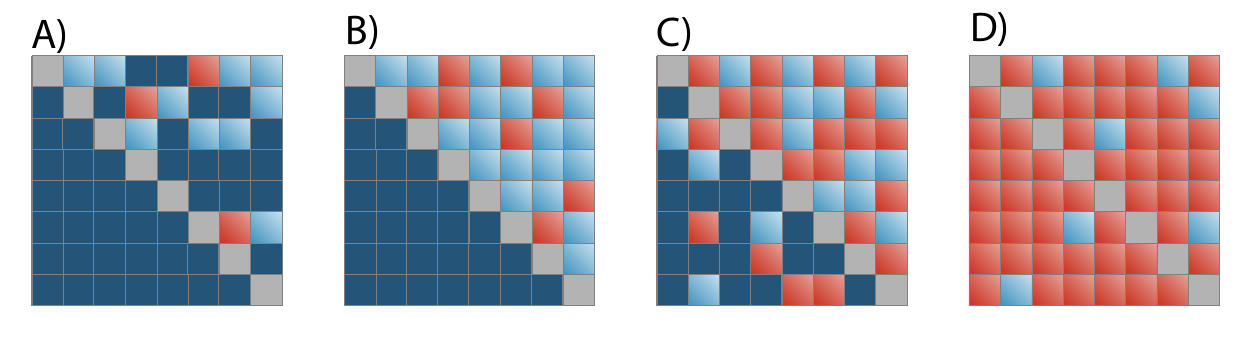}
\caption{Schematic representation of a transition from a centralized to a decentralized information flow scenario. If, for any given pair $(x,y)$, $T^S_{x,y} \sim T_{x,y}$, all existent dynamical driving is net driving, i.e., subsystems present a highly hierarchical structure. In this scenario, if a subsystem dominates another one, the former is not dominated by the latter. This is well illustrated in panels (a) and (b). Note however that in (a), only a few subsystems play an active (dynamical) role; whereas in (b) the situation has reached a perfectly hierarchical structure. Indeed, in this idealized situation the net transfer entropy reaches its maximum: any further addition in terms of dynamical driving will decrease the amount of net transfer entropy (as in panel (c)). Furthermore, (b) and (c) illustrate that there exists a tipping point beyond which the event has necessarily gone global. The extreme case where every subsystem exerts some amount of dynamical driving results in a ``null driving'' scenario, panel (d). In this schematic representation the color scales goes from dark blue to red, i.e. zero to maximum transfer entropy, respectively.}
\label{fig4}
\end{figure*}

In general, the evolving effective networks reveal a transition from a scenario with directed, hierarchical causal relationships to a symmetric {\em though rather fluctuating} networks where information is flowing symmetrically among all subunits. If information flows mainly in one direction (that is, if the sub-systems are arranged in a highly hierarchical structure) a subunit dominates another, with no or little information flowing in the opposite direction. In this situation, a convenient manipulation of the matrix $\left(T\to T^\dagger\right)$ based on a ranking and reordering of the elements according to their directionality index yields an upper triangular matrix (see Materials and Methods). The transition between such hierarchical or centralized driving to a symmetric scenario can be clearly identified monitoring the ratio $\theta=T^\dagger_l/T^\dagger_u$ between the sum all elements of $T^\dagger$ in the lower triangle and the same quantity evaluated in the upper triangle. As schematically illustrated in Figure~\ref{fig4}, in a regime of perfect directed driving all the elements below the diagonal are zeros, i.e., $\theta\approx 0$. In the opposite situation (i.e. the perfectly symmetric regime) the values below and above the diagonal are comparable, i.e. $\theta \approx 1$. The quantity $\theta$ can thus be considered as a suitable order parameter to characterize this order-disorder transition,  thus helping to identify and differentiate communication patterns across the subunits of a system. 

Figure~\ref{fig5} shows the behavior of the parameter $\theta$ as a function of time in our five datasets. In all the cases we initially observe a highly asymmetric effective network, where a few subunits have a dominant directional coupling to the rest of the system and $\theta\ll 1$. As the systems approach the onset date of the collective event, the quantity $T^\dagger_l/T^\dagger_u$ undergoes a quick transition to $\theta\approx 1$ identifying a regime in which the couplings indicate the existence of collective phenomena where all subunits are mutually affecting each other. We see that in four out of the five datasets the system has a clear order-disorder transition occurring in the proximity of the collective event. Interestingly, in the case of the Brazilian protests the measure significantly increases before the main event (June $17^{th}$). Such behavior probably results from the effects of small precursor protests taking place from June $6^{th}$ onwards. The same behavior is observed in the Higgs boson dataset, given the existing rumors triggered after July $2^{nd}$. Once more, the Google dataset behaves in a completely different way, never showing a clear signature of a collective regime for the couplings network. In the SI material we report the same analysis using the randomized signal for both the 15M and the Brazil events, and we observe no order-disorder transition.
 
\section{Discussion}
The mapping of influence networks using an information theoretic approach offers a new lens to analyze the emergence of collective phenomena. Through this lens, we can uncover the effective network of information flow between spatially defined sub-units of the social system and study the structural changes of the network connectivity pattern as the system goes through different collective states. In addition, the effective network lends itself to further analysis that can lead to the identification of structural hubs, coordinated communities, and geographical sub-units that may have recurrent roles in the onset of social phenomena. The methodology we present here can therefore be used to gain new insights on the structural and functional relations occurring in large-scale structured populations, eventually leading to the identification of metrics that might be used for the definition of precursors of large-scale social events.

Additionally, the methodology presented here opens interesting paths to advance in the analysis of social phenomena and the identification of generative mechanisms; however, this advance should not be conflated with the possibility of forecasting the emergence of social events. The evidence we discuss is agnostic with regard to the predictive potential of online networks and micro-blogging platforms. A real predictive approach cannot be disentangled from an automatic selection of the relevant discussion topics. Our analyses use datasets that were already zooming into the right conversation domain and monitoring specific keywords/hashtags in the Twitter stream. We believe, however, that the general methodological framework we put forward is a first step towards a better understanding of the temporal and spatial signatures of large-scale social events. This advancement might eventually inform the development of tools that can help us anticipate the emergence of macroscopic phenomena. In the meantime, our method offers a valuable resource to analyze how information-driven transitions unfold in socially relevant contexts.

\section{materials}
{\bf Data.} The first dataset focuses on the Spanish 15M movement, which emerged in 2011~\cite{borge2011structural,gonzalez2011dynamics}. The data cover a dormant period of low micro-blogging activity that is followed by an explosive phase in which the movement gained the attention of the general public and was widely covered by traditional media sources (see Figure~\ref{fig1}). The second dataset contains over 2.5 million geolocalized tweets associated to the {\em Outono Brasileiro} (``Brazilian Autumn''), a set of political protests that emerged in Brazil in June 2013. Similarly to the Spanish case, the Brazilian data include an initial phase of low activity followed by a gradual escalation towards the high volumes of general attention that accompanied the street protests. The third dataset tracks communication on the discovery of the Higgs boson before and after it was officially announced to the press in July of 2013; this dataset has been used before to assess how rumors spread through online social networks \cite{de2013anatomy}. The fourth dataset contains messages related to the release of a Hollywood blockbuster, announced months prior to its premiere to stir momentum amongst the fan base.  Finally, we also consider a dataset tracking communication on the acquisition of Motorola by Google, which came as sudden and unexpected news and immediately triggered a high volume of public attention. 

Spanish Twitter activity is spatially coarse-grained according to the list of metropolitan areas defined by the European Spatial Planning Observation Network~\footnote{See http://www.espon.eu. Accessed Apr. $16^{th}$, $2014$.}. This process yields $56$ aggregated time series: each of them corresponds to a different geographical area. In addition, there is an extra signal that accounts for any activity not included in those areas, i.e. the system is made up of $N = 57$ components. The data from Brazil are aggregated in 97 basins, which correspond roughly to metropolitan areas~\cite{Balcan_2010,Balcan2009-BMCMed}. The data tracking rumors about the Higgs boson are aggregated at the country level, including only the $N = 61$ most active around this topic. Finally, the Motorola-Google and the blockbuster data are classified in 52 U.S. areas: 50 states, plus Washington D.C and Puerto Rico. 

{\bf Order-Disorder Transition.} In real datasets the transition between the different scenarios can be visually inspected with a convenient sorting of the rows and columns of the $T_{x,y}$ matrix. We do so in Figure 5 of the main text, ranking each subunit of the system. The rank for a subunit $x$ is assigned according to the number of times $x$ it is dominant over the rest of the subunits. Once the ranking is settled, any $T_{x,y} < \frac{1}{2}T^{max}_{x,y}$ is set to 0 to improve the visual understanding of the figure. We then obtain a transformed matrix, i.e. $T_{x,y} \rightarrow T^\dagger_{x,y}$. Beyond visualization, the sorted matrix gives room to a monitoring measure $\theta=\frac{\sum_{x>y}T^\dagger_{x,y}}{\sum_{x<y}T^\dagger_{x,y}}=\frac{T^\dagger_l}{T^\dagger_u}$ (i.e., the ratio between the sums of all the matrix's elements in the lower and upper triangles) which provides a quantification of the state in which the system is (as explained in the main text).

\begin{figure*}
\begin{center}
\includegraphics[width=\textwidth]{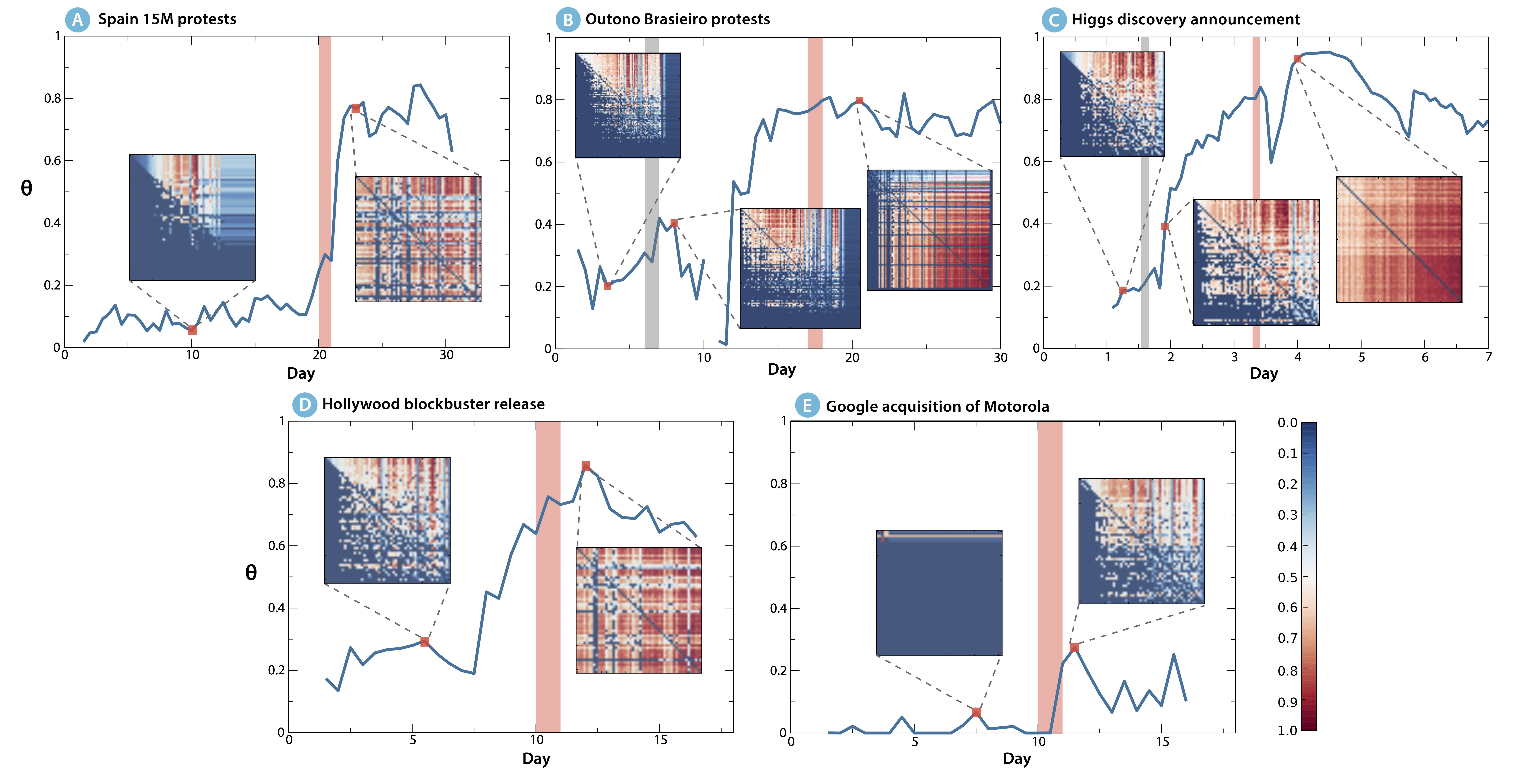}
\end{center}
\caption{\label{fig5} {\bf Order Parameter $\theta$ as a function of time for the five events analyzed.} The figure represents the behavior of the ratio $\theta=T^\dagger_l/T^\dagger_u$ characterizing the order/disorder of the effective connectivity matrix as a function of time (note a point missing in the Brazilian dataset due to a data blackout between days $10$ and $11$). For each dataset two or three matrices $T^\dagger$ are plotted considering one or two times before and one after the main event (signaled with a red vertical bar). A clear transition from a hierarchical directed to a distributed symmetrical scenario is observed for the events A, B, C and D. The Google dataset, depicted in panel E, behaves differently by not showing the same evidence of transition effects.}
\end{figure*}

\section*{\small{Acknowledgments}}
\small{AA acknowledges the support of the European Union MULTIPLEX 317532, the Spanish Ministry of Science and Innovation FIS2012-38266-C02-01, and partial financial support from the ICREA Academia and the James S. McDonnell Foundation. YM acknowledges support from MINECO through Grant FIS2011-25167; Comunidad de Aragón (Spain) through a grant to the group FENOL, and by the EC FET-Proactive Project MULTIPLEX (grant 317532). For the analysis of data outside of the United States of America AV and NP acknowledge the Intelligence Advanced Research Projects Activity (IARPA) via Department of Interior National Business Center (DoI/NBC) contract number D12PC00285. The views and conclusions contained herein are those of the authors and should not be interpreted as necessarily representing the official policies or endorsements, either expressed or implied, of IARPA, DoI/NBE, or the United States Government. The funders had no role in study design, data collection and analysis, decision to publish, or preparation of the manuscript. We thank D. Allen, R. Compton and T-C Lu at HRL Laboratories LLC for assistance with the Brazilian dataset and useful discussions; we also thank A. Lima for sharing the Higgs boson data.}



\begin{center}
The dynamic of information-driven coordination phenomena: a transfer entropy analysis\\
\textsc{Supplementary Information}
\end{center}

\noindent{Javier Borge-Holthoefer,$^{\ast}$ Nicola Perra,$^{\ast}$ Bruno Gon\c{c}alves, Sandra Gonz\'alez-Bail\'on, Alex Arenas, Yamir Moreno,$^{\ast}$ Alessandro Vespignani$^{\ast}$}

\noindent{\normalsize{$^\ast$To whom correspondence should be addressed: jborge@qf.org.qa, n.perra@neu.edu, yamir.moreno@gmail.com, a.vespignani@neu.edu}}




\renewcommand\thefigure{S\arabic{figure}}
\setcounter{figure}{0}
\setcounter{section}{0}
 
\section{Data, Context and Chronology of the Events Analyzed}

We considered five different events: the Spanish 15M protests, the 'Outono Brasileiro' (Brazilian autumn) movement, the announcement of the Higgs boson discovery, the release of a Hollywood blockbuster movie (Batman ``The Dark Knight Rises''), and the acquisition of Motorola by Google. In this section we report details concerning these events and the associated Twitter data sets.


\subsection{The 15M Protests (May 2011)}

These protests emerged in Spain in the aftermath of the so-called Arab Spring. A grassroots social movement, later called the 'Indignados' (``the outraged''), it emerged from online communication amongst a decentralized network of citizens and civic associations. Online networks (blogs, Facebook, Twitter) were used to spread a call for action for May 15, 2011. The main drivers of the protests were spending cuts and policy reactions to the economic crisis. Massive demonstrations took place on May 15 in several major cities around Spain, many of them resulting in camp sites in main city squares that remained active for weeks. Mainstream media didn't cover the movement until it reached the streets. As a consequence, most communication and broadcasting announcing and discussing the mobilizations took place through online channels. Social media networks (in particular, Twitter) served a crucial role in the coordination of the protests and the management of camp logistics \cite{castells2013networks,gerbaudo2012tweets}.

The Twitter data for the Spanish 15M movement were harvested by a startup company ({\em Cierzo Ltd.}) for a period spanning from April 25 to May 25, 2011. The main demonstrations in Spain took place on May 15 and onwards, thus our analysis covers a brewing period with low activity rates (up to May 15, day 20 in the Figures) plus an ``explosive'' phase beyond that date, in which the phenomenon reached general public and was widely covered by traditional mass media, see Figure 1 in the Main Text. Scraps on Twitter servers yielded $581,749$ messages. 

\subsection{'Outono Brasileiro' Protests (June 2013)}

More recently, massive protests filled the streets of several Brazilian cities. The triggering factor was the rising prices of public transportation, but on the background loomed long-standing discontent with inequality, the government economic policies, and the provision of social services. Social media played again an instrumental role in the coordination of large-scale mobilization and the swift diffusion of information: images documenting the often brutal police reaction to the protests boosted mobilization and brought more people to the streets of more cities and municipalities. The protests, often dubbed as 'Outono Brasileiro' (``Brazilian autumn''), resulted in Brazilian President Dilma Rouseff announcing, in June 21, measures to improve the management of public transport along with other social services. This prime-time televised address, however, did not placate citizens dissatisfaction, who continued staging protests in subsequent days.

The dataset regarding such event has been obtained using the PowerTrack tool that provides $100\%$ coverage for a set of specified keywords (see table~\ref{table1}). For our analysis we considered just the tweets sent in the month of June 2013 ($2,670,933$ tweets). Indeed, the first large scale protest, often associated with the escalation of the protests, took place on June 17th, with remarkable (though smaller) precursors on the 6th and 13th. As in the case of the Spanish movement we considered a brewing period with low activity rates plus the ``explosive'' phase beyond the date of the first massive street protest.

\begin{table}
\begin{tabular}{l|l|l|l|l}
AnonymousBrasil & boicot & cacerolada &cacerolazo & huelga \\
marcha & marchado &marcham & marchamos & marche  \\
march\'e & marcho &Passeata & protesta &protestam \\
protestar\'as &protestarem &protestarmos & proteste &protestemos\\
protesten &protesto &protest\'o & concentraci\'on & reforma\\
greve & rali & manifesta\c c\~ao & manifestantes &corrup\c c\~ao\\
\end{tabular}
\caption{List of keywords used to find tweets related to the 'Outono Brasileiro'}
\label{table1}
\end{table}

\subsection{The Higgs Boson Discovery Announcement (July 2012)}

In July 4 2012, a team of scientists based at CERN presented results that indicated the existence of a new particle, compatible with the Higgs boson (the existence of which had first been hypothesized in 1964). Mainstream news media covered the discovery after the announcement, but during the days preceding it there were already rumors of its discovery circulating through social media \cite{de2013anatomy}. The messages we analyze were collected using Twitter's publicly available API between July 1 and July 7 using a list of relevant keywords (i.e. lhc, cern, boson, higgs). In total, the data set contains 985,590 tweets.

\subsection{The Hollywood Movie Release (July 2012)}

The Dark Knight Rises is the third installment of the Batman trilogy (started in 2005 with the release of Batman Begins and followed up in 2008 with The Dark Knight). It was premiered in New York on July 16 2012, and released in several English-speaking countries a few days later. The promotional campaign included so-called viral marketing through social media. The film was nominated to several prestigious awards, and grossed over a billion dollars in the box office.

The dataset includes $130,529$ tweets between July 6th and July 21st that include the words ``batman'', ``darkknight'' or ``darkknightrises''. The tweets are obtained from the Twitter Gardenhose (a 10\% random sample of the entire Twitter traffic).
  
\subsection{The Google-Motorola Acquisition (August 2011)}

On August 15, Google announced a relatively unexpected agreement to acquire the mobile company Motorola.  The move was a strategic attempt to strengthen GoogleÕs patent portfolio in a context where legal battles over patents is increasingly shaping the mobile industry and the telecommunications environment.

The dataset contains $10,890$ tweets between August 5th and August 20th, 2011. In order to minimize the noise, we considered just tweets containing both ``google'' and ``motorola''. Also in this case, the tweets are obtained from the Twitter Gardenhose.

\section{Methods used in the analysis}

In this section we detail how the Twitter time series are constructed.

\subsection{Data spatial aggregation}

With activity information at hand, a possible way to represent information is to assign a time series to individual Twitter users. This however has important drawbacks: activity may be too sparse to build a significant series; it may be rather difficult to detect general, meaningful trends when studying series interaction; finally, one needs to take into account computational costs.

We have chosen to coarse-grain the data from a geographical point of view. We believe this has several advantages, among which: (i) the number of significant units will be relatively low, easing our capacity to analyze the results; and (ii) geographical units (metropolitan areas, states) stand as useful entities in social research regarding personal interactions, political activity, economic transactions, etc. See~\cite{conover2013geospatial,jones2013yahtzee} as recent examples of the geographical approach.

For the 15M case, geographical information was collected for each user involved in the protests, thereafter tweets were assigned their author's location. Spanish Twitter activity is spatially coarse-grained according to the list of metropolitan areas defined by the European Spatial Planning Observation Network (http://www.espon.eu). This process yields 56 aggregate time series, each corresponding to a geographical area, plus an extra signal which accounts for any activity not included in the previous definition. Thus, the system is made up of $N=57$ components. Time-stamps have been modified when necessary (Santa Cruz de Tenerife, Orotava and Palmas de Gran Canaria) to a common time frame. The pre-defined metropolitan areas account for over half Spain's total population.

The brazilian tweets have been instead aggregated at the level of $N=97$ basins centered around major transportation hubs. These geographical units, that correspond to census areas surrounding large cities, have been defined aggregating population cells of $15 \times 15$ minutes of arc \footnote{ This corresponds to an area of each cell approximately equivalent to a rectangle of $25 \times 25 \, kms$ along the Equator}, from the "Gridded Population of the World" and the "Global Urban-Rural Mapping" projects~\cite{pop_dat,pop_dat2}, to the closest airport that satisfies the following two conditions: $(i)$ Each cell is assigned to the closest airport within the same country.
And $(ii)$, the distance between the airport and the cell cannot be longer than $200 \, kms$. This cutoff naturally emerges from the distribution of distances between cells and closest airports. See refs.~\cite{balcan09-2,balcan09-1} for details. Moreover, having access to $100\%$ of the entire signal on Twitter associated with at least one word listed in Table~\ref{table1}, we considered just tweets with live GPS coordinates. 

Tweets around the Higgs boson discovery were aggregated at the country level. The original dataset contained tweets from over 200 countries, but these have been thresholded to retain only those countries with more than 500 tweets over the topic of interest, for the whole week. This entails that only 61 countries are present in the analysis in the main text. The details regarding the location technique can be found in \cite{de2013anatomy}.

The remanning two datasets have been aggregated in 52 areas --50 U.S. states, plus Washington D.C and Puerto Rico. In particular, 
the geographical information of tweets in this case has been gathered either from live GPS locations, or mining the so called ``self-reported location". In general this field is filled freely by the users that can report their location at different levels. i.e. NYC, California, CA, USA etc.. Some fraction of the reported locations are jokes, i.e. moon, mars, behind you etc. We parsed these fields trying to match a country, state, or city name. In Figure~\ref{geolocation} we report the flow chart of the algorithm used.
Interestingly, the method is able to find a match of $40-50\%$ the total number of tweets.

\begin{figure}[h!]
\centering
\includegraphics[width=0.7\linewidth,clip=0]{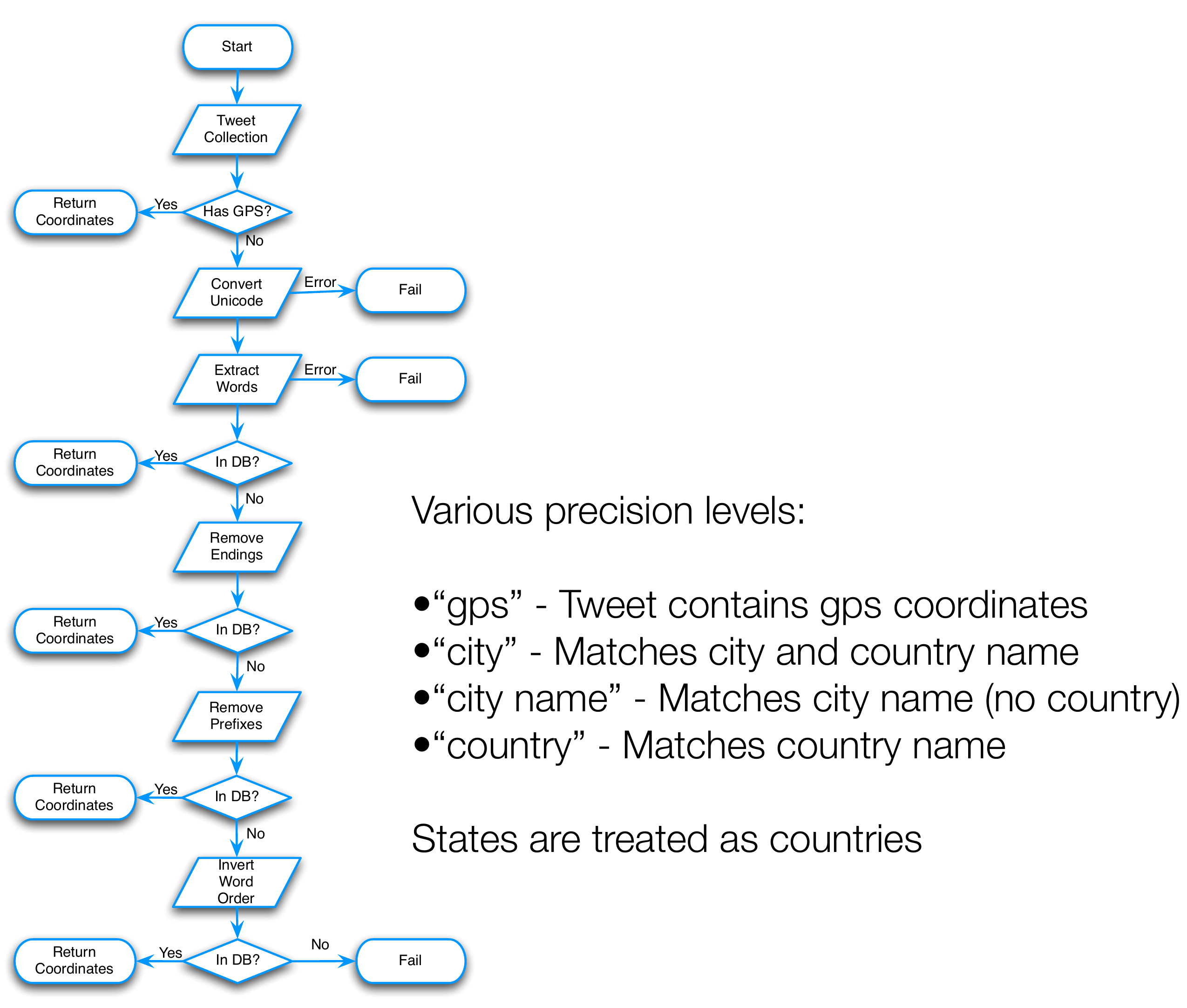}
\caption{Schematic representation of the algorithm used to gather geographical coordinates of the Hollywood movie release and the Google-Motorola acquisition datasets. In the figure DB stands for database. We used the ``GeoNames"  database~\cite{GeoName}.}
\label{geolocation}
\end{figure}

In Section~\ref{geosens} we show the results obtained considering different geographical aggregations. Namely, we also considered $N=16$ Spanish communities (one of them was left out because no messages were collected from there), $N=27$ Brazilian states, and $N=9$ regions in the USA defined by the American Census Bureau~\cite{usa_census}.   

\subsection{Data temporal aggregation}

The definition of the temporal aggregation of Twitter data is particular important in our approach. Indeed, we want to determine the characteristic time scale at which the driving between series is most evident. The data comes with temporal resolution down to a second. However, such level of resolution is excessive to detect dynamical trends among series. We considered different sampling rates $\Delta \tau$ spanning from $1$ to $120$ minutes. Although arbitrary, these range of temporal aggregations account for the fluidity of Twitter's discussion as well as the limited attention time span of users. In Section \ref{cts} we discuss the ideas that allow us to determine, among the temporal aggregation schemes, the optimal one.

\subsection{Symbolic Transfer Entropy}

Closely related to other measures, such as mutual information \cite{shannon48}, Granger causality \cite{granger1969investigating} and transfer entropy \cite{schreiber2000measuring}, Symbolic Transfer Entropy (STE) \cite{staniek2008symbolic} provides a solid method to detect and quantify the strength and direction of couplings between components of dynamical systems. The symbolic approach, on the other hand, links STE to order patterns and symbolic dynamics \cite{bandt2002permutation,schinkel2007order} as a means to successfully analyze time series which may be noisy, short and/or non-stationary.

Once spatial and temporal aggregation schemes are fixed, we proceed to measure STE as a way to quantify the coupling among series. Note that such series span long times, $L$, of several days or even a month. Also, activity during these days is changing due to offline events happening outside the Twittersphere. Thus, STE is not measured over time series taken as a whole, but over sliding windows of length $\omega \ll L$ (which is indeed a standard way to proceed in neuroscience). To obtain a finer analysis, these windows advance at a slow pace of only 30 minutes. In practice, this means that the first window spans the interval $[0,\omega]$; the second one $[30, \omega+30]$ (in minutes), and so on. Window width $\omega$, admittedly, is the first parameter that will affect the measurements output, and we will discuss its effects later.

Given a window of width $\omega$, the resulting series are transformed into symbol sequences as described in \cite{staniek2008symbolic}, for which an embedding dimension $3 \le m \le 7$ \cite{bandt2002permutation} must be chosen (see also Section~\ref{dism} for further details). Let us consider a simple example of how this works. Imagine we have a signal 
\begin{equation}
x=\{120, 74, 203, 167, 92, 148, 174, 47\}
\end{equation} (let us ignore sliding windows by now). We shall transform this series into symbol series. For simplicity, let us suppose that the embedding dimension $m=3$. This quantity determines the amount of symbols that can possibly exist, which is $m! = 6$ in our case. See Figure \ref{m3} as an illustration of the possible symbols that can be obtained.

\begin{figure}[h!]
\centering
\includegraphics[width=1.0\linewidth,clip=0]{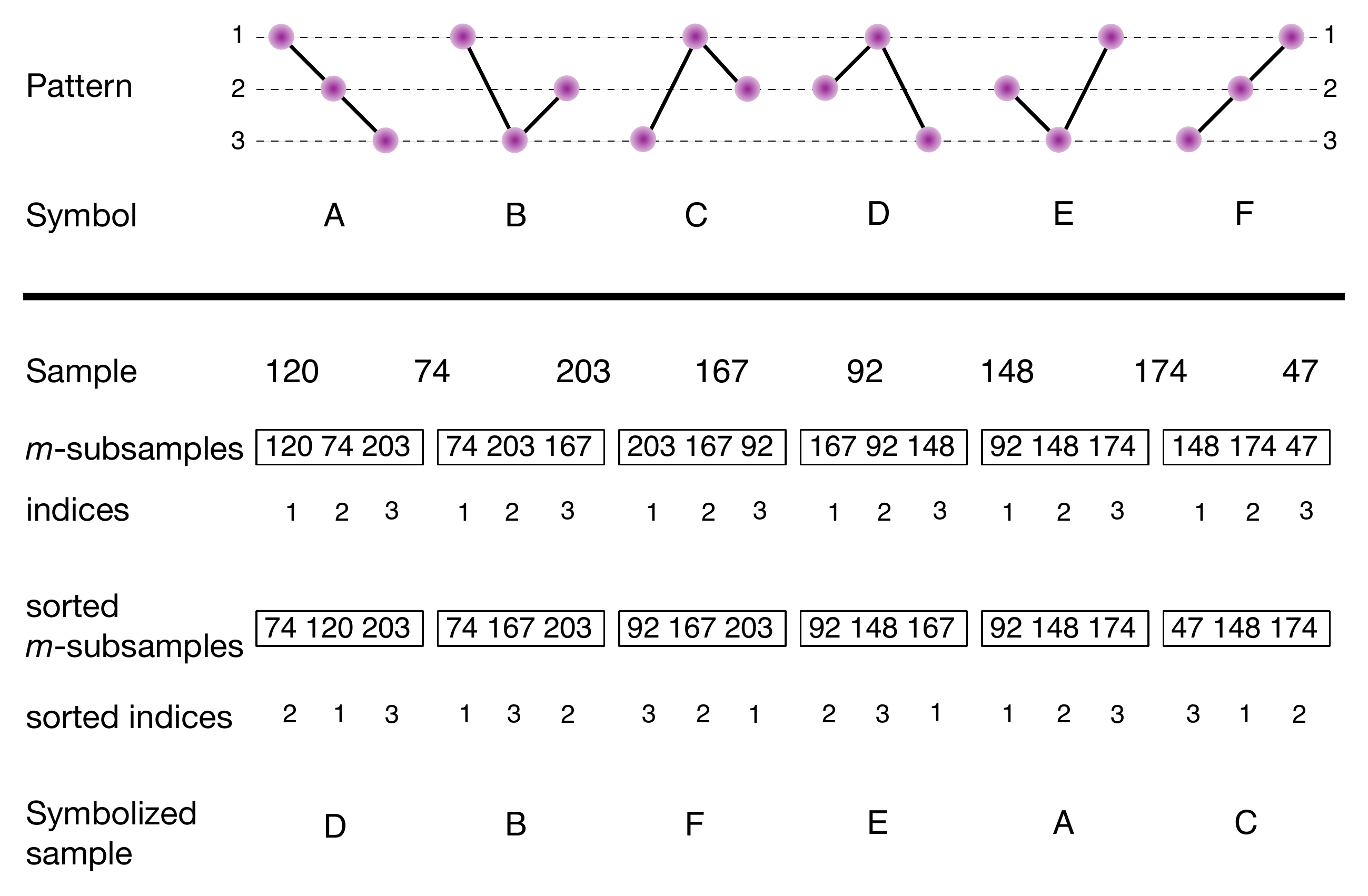}
\caption{Sample order pattern for $m = 3$. If we neglect ties, the number of possible patterns is 3! (see \cite{staniek2008symbolic} to check how ties are dealt with).}
\label{m3}
\end{figure}

The first step to transform $x$ into symbol sequences is to sort their subchains of length $m$ in increasing order. So, we take the first three elements of $x$ and sort them, which leaves us with $\{74, 120, 203\}$. We have kept track of these values' indices, such that the sequence now looks like $\{2, 1, 3\}$. According to Figure \ref{m3} (top), this first subchain maps to the symbol D.

From this scheme, we just need to advance one value at a time: the next subchain to consider is $\{74, 203, 167\}$. Its sorted version is $\{74, 167, 203\}$, which corresponds to $\{1, 3, 2\}$, and maps to B. The whole process for the $x$ signal looks like Figure \ref{m3} (center), and their sorted indices lead to Figure \ref{m3} (bottom), rendering a symbol sequence $\hat x = \{D, B, F, E, A, C\}$. With a similar procedure, other series $y$ are transformed into $\hat y$. Given these symbol sequences $\{\hat x\}$ and $\{\hat y\}$, STE between a pair of signals $(x,y)$ is defined as

\begin{equation}
T_{yx} = \sum p(\hat x_{i+\delta}, \hat x_{i}, \hat y_{i}) \log \frac{ p(\hat x_{i+\delta} \mid \hat x_{i}, \hat y_{i})} {p(\hat x_{i+\delta} \mid \hat x_{i})} 
\end{equation}
where the sum runs over all symbols and $\delta$ denotes a time step. 

A few facts need to be highlighted at this point:

\begin{enumerate}
\item A signal with an original length of $n$ points is reduced, through symbolization, to a new string with $n-m+1$ symbols.
\item A way to interpret the meaning of $m$ is to think of it as the amount of ``expressiveness'' it allows to the original series. That is, if $m$ is low, a rich signal (one with many changes in it) is reduced to a small amount of possible symbols. This is of utmost importance to understand why we have chosen a relatively high $m$ to work with (see Section \ref{dism}).
\item All measurements in the present work have been performed using $\delta = 1$. This implies that we are measuring the capacity of a signal to predict the {\em immediate} future of another signal, i.e. just one symbol ahead.
\end{enumerate}

This measurement for each pair of signals is encoded in a matrix $T$ which contains pairwise information about how each system's component dominates others (or is dominated). Note that $T$ is an asymmetric matrix. That means that a certain source $x$ can exert some driving on $y$, and at the same time $y$ might exert some driving on $x$. To see how information flow balances, the matrix $T^{S}_{xy} = -T^{S}_{yx} = T_{xy} - T_{yx}$ is built.

%

\subsection{Defining the characteristic time scale of the events}
\label{cts}
 
Given a certain dataset, we do not have any prior knowledge to define the correct timescale $\Delta t$ at which time series should be aggregated. We do know, however, that activity around civil protests (and in general around any event that involves collective action) are typically far from being stationary or periodic which intuitively points at the fact that there might not exist a single time scale for the whole dataset.

Let's consider for now a fixed temporal resolution $\Delta t$ and bin the Twitter activity such that the first point in the time series will contain any activity that happened between $[0, \Delta t)$; the second point will contain data from $[\Delta t, 2\Delta t)$, and so on. For example, a 30 day dataset, sampled at $\Delta t = 60$ minutes will render a set of time series of length $L = 30 \times 24 = 720$ points.

As mentioned above, the STE is measured using signals in windows of width $\omega$, spanning from $[t-\omega, t)$. In order to define the optimal $\Delta t$ we evaluate which temporal resolution provides the best possible information flow among units, i.e.  the optimal $\Delta t$ is the one containing more transfer entropy $s_{\Delta t}=\sum_{x,y} T^{\Delta t}_{xy}$. Operatively, the best timescale is defined as $\{ \Delta t: \max_{i}(s^{i}_{\Delta t}) \}$. We considered as possible candidates all values $\Delta t$ from $1$ to $120$ minutes, with increases of 5 to 15 minutes.

It is important to notice that the maximization of STE at each sliding window may result in different optimal $\Delta t$: our proposal lets the data inform about the time scale at which events are best described. Such fact, due to changes in Twitter activity, is illustrated in Figure 2 in the main text. 

\subsection{Defining the information flow of the events}

One may further scrutinize the temporal evolution of the amount of STE each component displays. Instead of studying pairwise information flows, in this case we focus on whether a geographical unit is on average driving others or is driven by others at each time step. Within each window of width $\omega$ at time $t$, we calculate the values of the net flow matrix for that window, or normalized directionality index (di), for each geographical unit $x$ at each time step defined as:
\[ x_{di} = \left\{ 
  \begin{array}{l l}
    \frac{\sum_{y}T^S_{xy}}{\max_{x}\sum_{y}T^S_{xy}} & \quad \text{if $\sum_{y}T^S_{xy}>0$ }\\
    \frac{\sum_{y}T^S_{xy}}{\min_{x}\sum_{y}T^S_{xy}} & \quad \text{if $\sum_{y}T^S_{xy}<0$ }
  \end{array} \right.\]
Thanks to the normalization $-1 \le x_{id} \le 1$ the largest value is associated to the geographical unit exercising the largest driving force to other units. Vice versa, the smallest value is associated to the geographical unit subject to the largest driving forces from other nodes. The result of this measurement --and the corresponding analysis-- can be found in Figures 3 in the main text. Note that each point in the panels of that figure condenses the results obtained for a window integrating information from the {\em past}, i.e. activity within $[t- \omega, t)$.

\section{Sensibility analysis of the parametrization}

Results in the main text have been obtained with 1-day (i.e., $\omega=1$) long sliding windows, and embedding dimension $m=5$, and using particular geographical units (metro areas in Spain, basins in Brazil, and states in USA). In this section, we present the results of various sensitivity analyses testing how each parameter influences the results. 

\subsection{The role of the sliding windows width $\omega$}

\begin{figure*}[t]
\centering
\includegraphics[width=0.7\linewidth,clip=0]{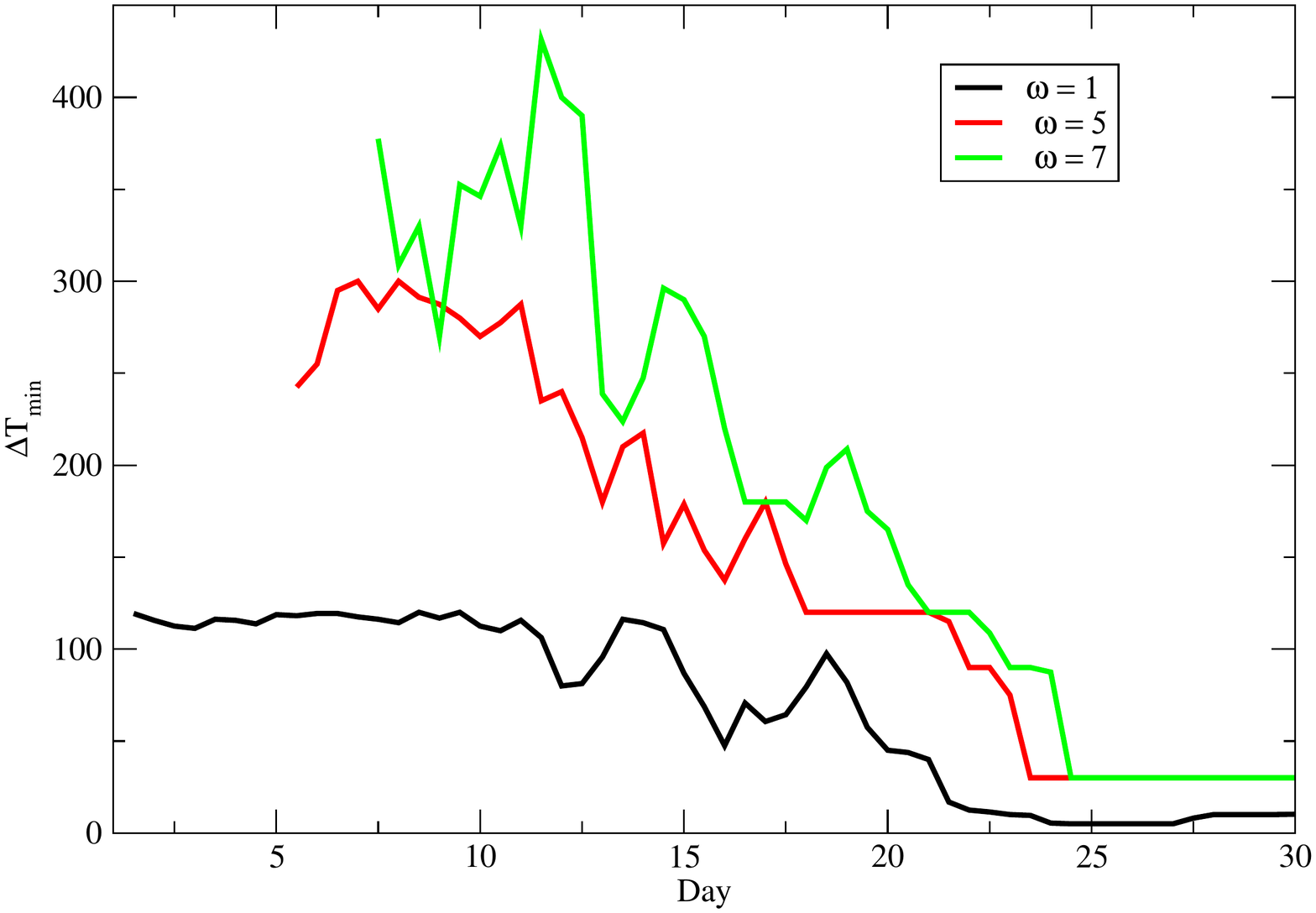}
\caption{Dependence with the sliding window size $\omega$  considering the Spanish 15M protest. Black, red and green lines describe $\omega=1,5,7$ respectively.}
\label{tscale_across_w}
\end{figure*}

Intuitively, if $\omega$ is set to a large value the capacity of the method to anticipate events will be reduced, because information emitted long before the present time is affecting the calculations. To check how larger $\omega$'s blur the results, we have reproduced the observations in the main text for $\omega=5$, and $\omega=7$ given a fixed $m=5$ considering the 15M dataset. Results are offered in Figure \ref{tscale_across_w}. It is clear that different window widths behave in the same way as the original one (1 day). Nevertheless, shorter $\omega$ yields a more abrupt transition close to the critical event. This observation is more evident studying the behavior of normalized directionality index. In Figure~\ref{bubble_same_m_different_w} we show this quantity at each time step for the case of the 15M protests in Spain. Similarly to figure 3 in the main text the size of each bubble is proportional to the logarithm of the activity on twitter. We notice that for $\omega=7$ and $\omega=5$ the system shows a change in the driving dynamics clearly before the 15 of May (red strip). Instead, for $\omega=1$ the transition form a scenario in which the large metropolitan areas are the major driving forces to a more homogenous and delocalized scenario happens during the unfolding of 15 of May.

\begin{figure*}[t]
\centering
\includegraphics[width=0.6\linewidth,clip=0]{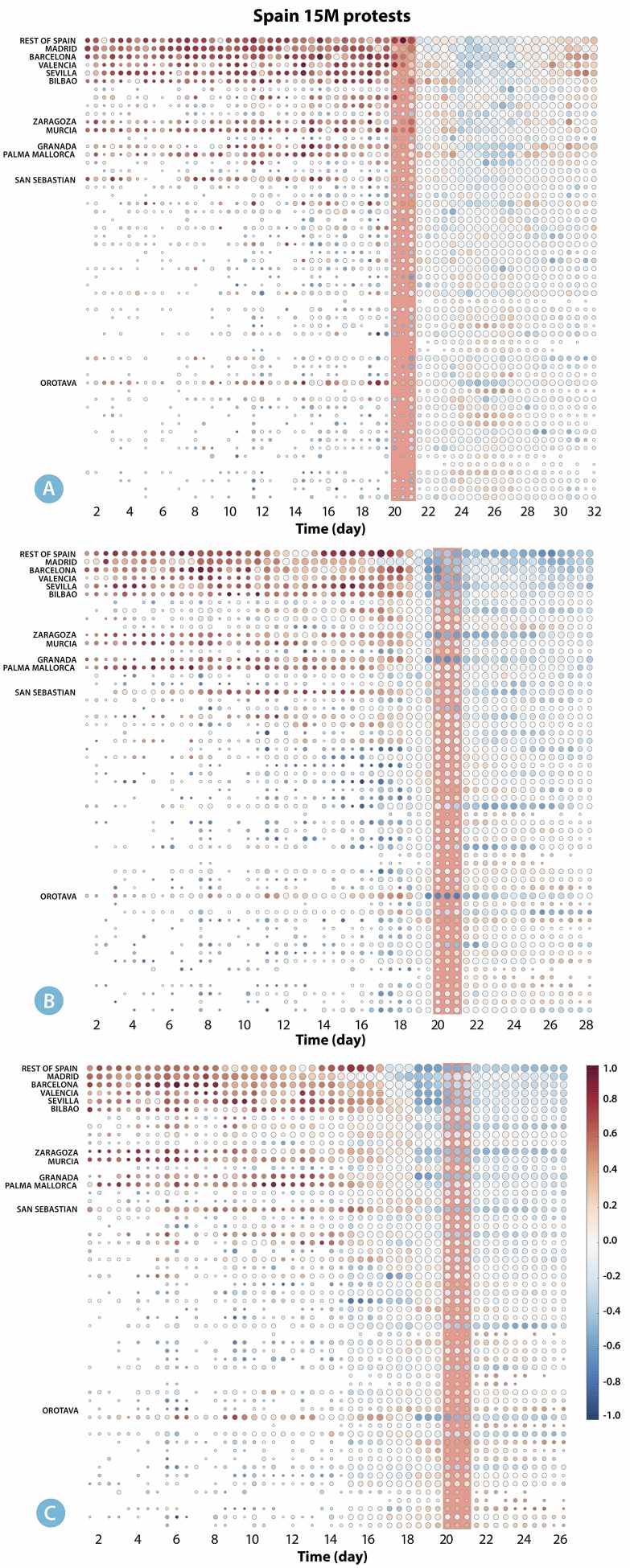}
\caption{Normalized directionality index for each geographical unit in the 15M dataset for different $\omega$. In panel A) we considered $\omega=1$, in panel B) $\omega=5$ and in panel C) $\omega=7$. For all the cases we set $m=5$. The red strip indicate the 15 of May}
\label{bubble_same_m_different_w}
\end{figure*}

\subsection{The role of the embedding dimension $m$}
\label{dism}
The embedding dimension $m$ determines how the information in the original time series will be transformed into symbols. The larger $m$, the larger is the collection of symbols onto which the values are mapped. Since the size of symbols grows like $m!$, it is clear that complex time series demand higher $m$ for a faithful mapping (i.e. one that collects the original complexity). On the other hand, overestimating $m$ adds unnecessary computational costs, because the final result won't change qualitatively. We address the problem of finding the minimal sufficient embedding dimension $m$, using the approach, called the false nearest neighbor method, proposed by Kennel {\em et al.} \cite{kennel1992determining}.\\
In practical terms the minimal value of $m$ is found studying the behavior of the nodes encoding the large majority of information. In the case of the 15M protests in Spain this corresponds to Madrid, which was a key spot for the grassroots movements. The minimal value of $m$ is sufficient to disentangle the signal from the dominant node and any other time series in the corresponding dataset will need the same or smaller $m$ to be faithfully mapped. Figure \ref{madrid} reflects these calculations and the strong conclusion is that for any $m \ge 5$ symbolization will have captured the original topology of the real data. Thus, all through the main text, and also in this document, results are reported for $m=5$, unless indicated otherwise.\\
In order to further study the effects of $m$ in Figure~\ref{diff_m_same_w} we plot the behavior of the normalized directionality index for $m=3,4,5$. As it is clear from the plot $m=5$ (panel C), is able to capture the transition from asymmetric to symmetric scenario in more details.

\begin{figure*}[t]
\centering
\includegraphics[width=0.7\linewidth,clip=0]{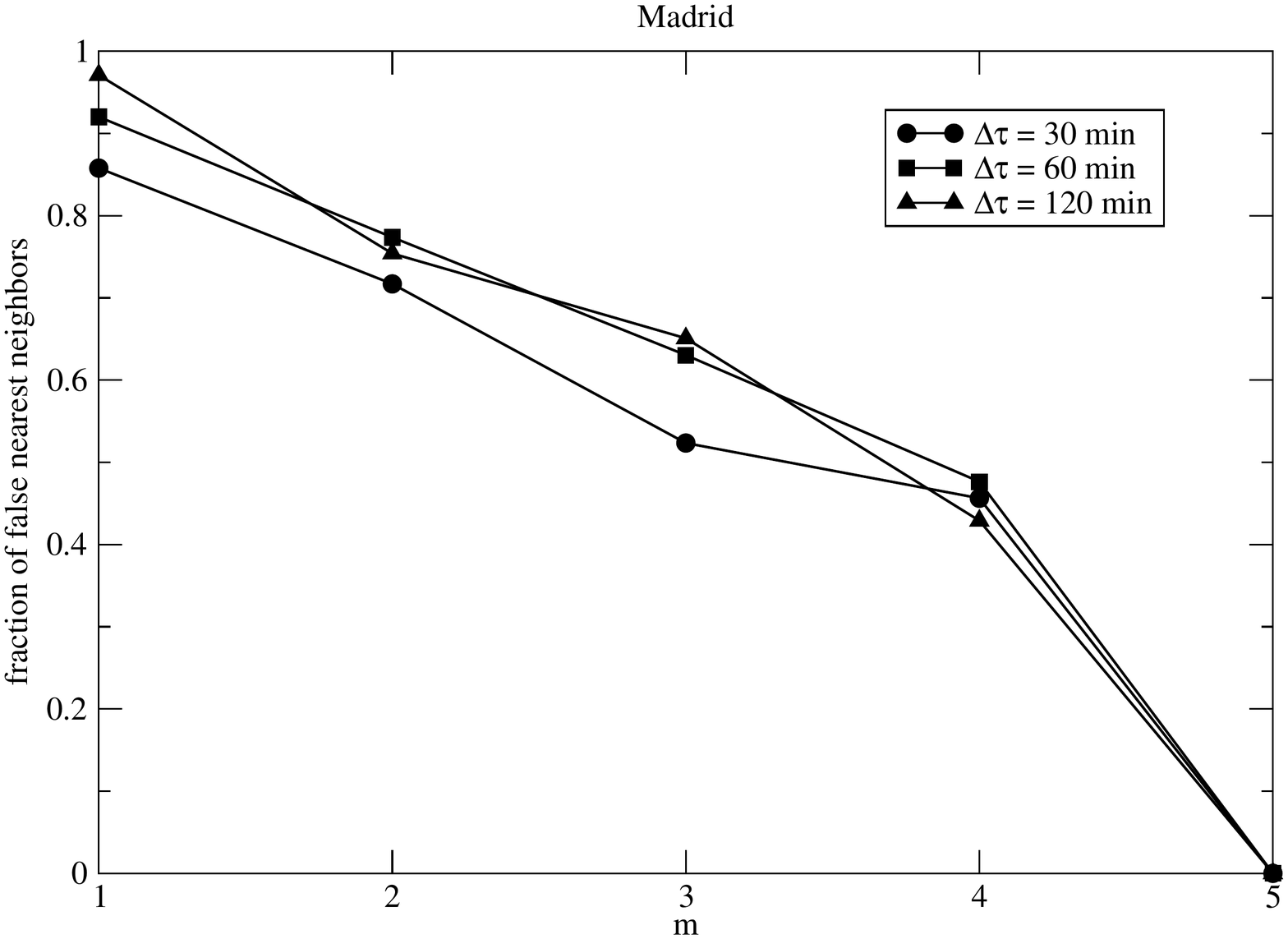}
\caption{Fraction of false nearest neighbors as a function of $m$ for the Spanish dataset and the Madrid time series.}
\label{madrid}
\end{figure*}

\begin{figure*}[t]
\centering
\includegraphics[width=0.6\linewidth,clip=0]{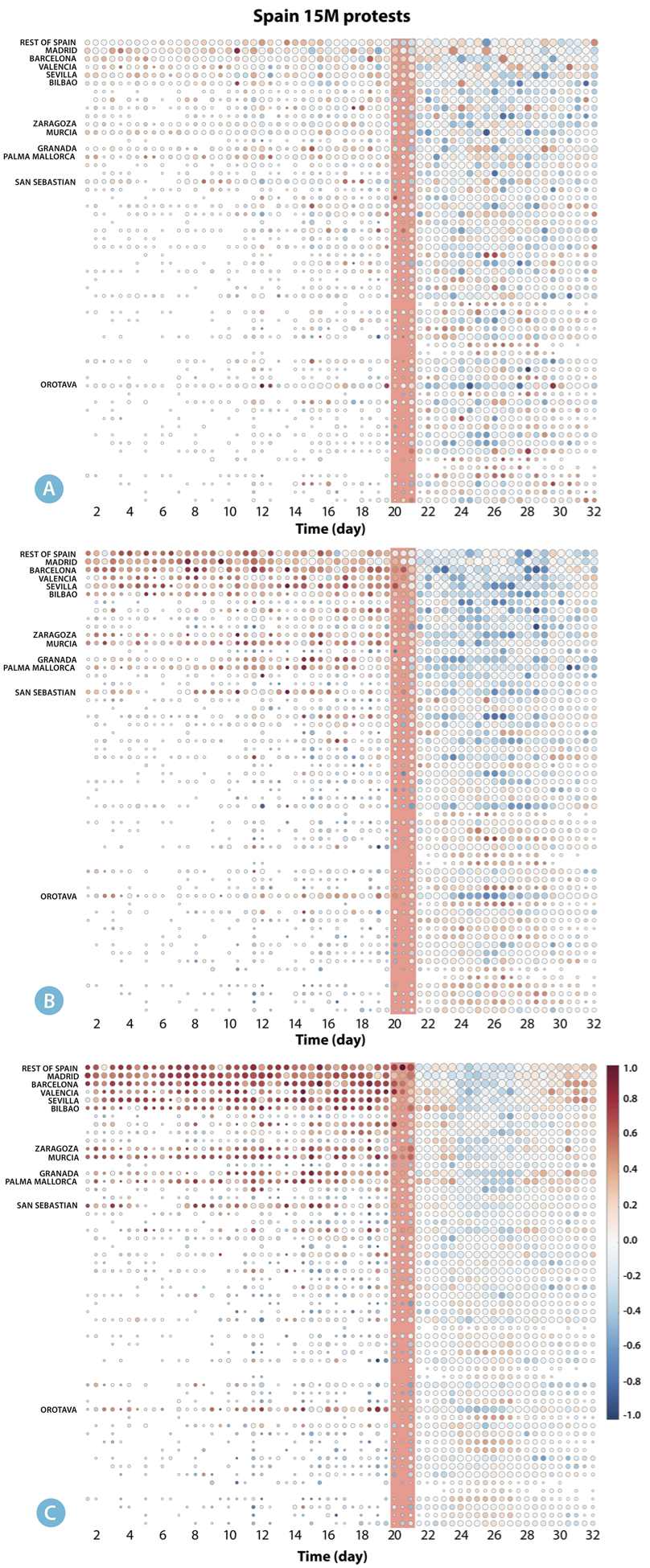}
\caption{Normalized directionality index for each geographical unit in the 15M dataset for different $m$. In panel A) we considered $m=3$, in panel B) $m=4$ and in panel C) $m=5$. For all the cases we set $\omega=1$. The red strip indicate the 15 of May}
\label{diff_m_same_w}
\end{figure*}

\subsection{Sensibility analysis of the partition}
\label{geosens}
As mentioned above, the Twitter signal could be represented in many ways --and we have chosen a geographical approach. In fact, an enormous range of settings are possible: from a simple bipartition of the activity stream to a complete breakdown where a single user is matched to a time series. It is fair then to state up-front that our decision is an arbitrary one, driven by the obvious fact that geography matters in the offline and online worlds.\\
Even within the geographical scheme, many options are available: spatial aggregation could be done at the neighborhood, city or county levels (for a finer resolution), or considering a coarser partition. For each dataset in the main text we have repeated the analysis for coarser geographical divisions. In the case of Spain's 15M movement, we have moved from the metropolitan areas to the autonomous community level (the Spanish 17 autonomous communities can be regarded as states, i.e. political entities at the regional level~\cite{spain}). Data from Brazil have been binned in 27 states~\cite{brazil}, in contrast with 97 basins in the main text. Finally, US data has been aggregated up to the ``divisions'' level (9 supra-state areas, as defined by the US Census Bureau~\cite{usa_census}).\\
Results for these alternative data partitioning can be seen in figures~\ref{fig2_si}, \ref{figure3_complete_different_geo} and~\ref{space_new}. Regarding the evolution of the time scales (figure~\ref{fig2_si}), we observe that the behavior qualitatively resembles the original one in the main text. A similar result is obtained for the information flow balance in figure~\ref{figure3_complete_different_geo}, where the occurrence of protests and demonstrations (15M and Outono Brasileiro) marks a change in the dominant pattern; the same can be said for the Google-Motorola case.\\

However, some differences appear in the $\theta - t$ plot (figure~\ref{space_new}). To start with, the Brazilian dataset and the Batman event deliver dense $T_{xy}$ matrices, i.e. its sorted counterpart $T'$ has many below-diagonal elements even at early times, indicating that no (or little) transition takes place: the system is decentralized from the very beginning. These differences demand some explanation.\\
First, it must be highlighted that our tip-over rationale (see subsection \ref{subsec:tipover}) is valid regardless the apparent contradiction: our claims are concerned with how the values in the $T_{xy}$ matrix are distributed, and as such it is an abstraction of what such matrix represents (be it cities, states or individuals, for that matter). Then, the apparent contradiction simply points at the fact that the lens through which we analyse the events {\em does} matter. Taking it to the extreme, a bipartition of the data, with two time series accounting for half of the activity each, would easily yield a fully symmetrical $T_{xy}$ matrix; in the opposite situation, a system comprising each user individually would render an (almost) empty matrix, given the fact that most people is not showing activity most of the time.\\
All in all, these results suggest that our proposal opens up exciting research questions: for instance, at which level of resolution should the system be observed to extract an optimal analysis out of it? We must keep in mind that other relevant events in Twitter do not have a geographical component; groups may be defined by religious beliefs, age strata, genre issues. It remains beyond the scope of this work to determine how to obtain optimal partitions that will render the correct conclusions. For the time being, we rely on commonsensical, predefined --rather than optimally detected-- entities (geographical, in this case) to make a case of our methods and rationale.

\begin{figure*}[t]
\centering
\includegraphics[width=0.8\linewidth,clip=0]{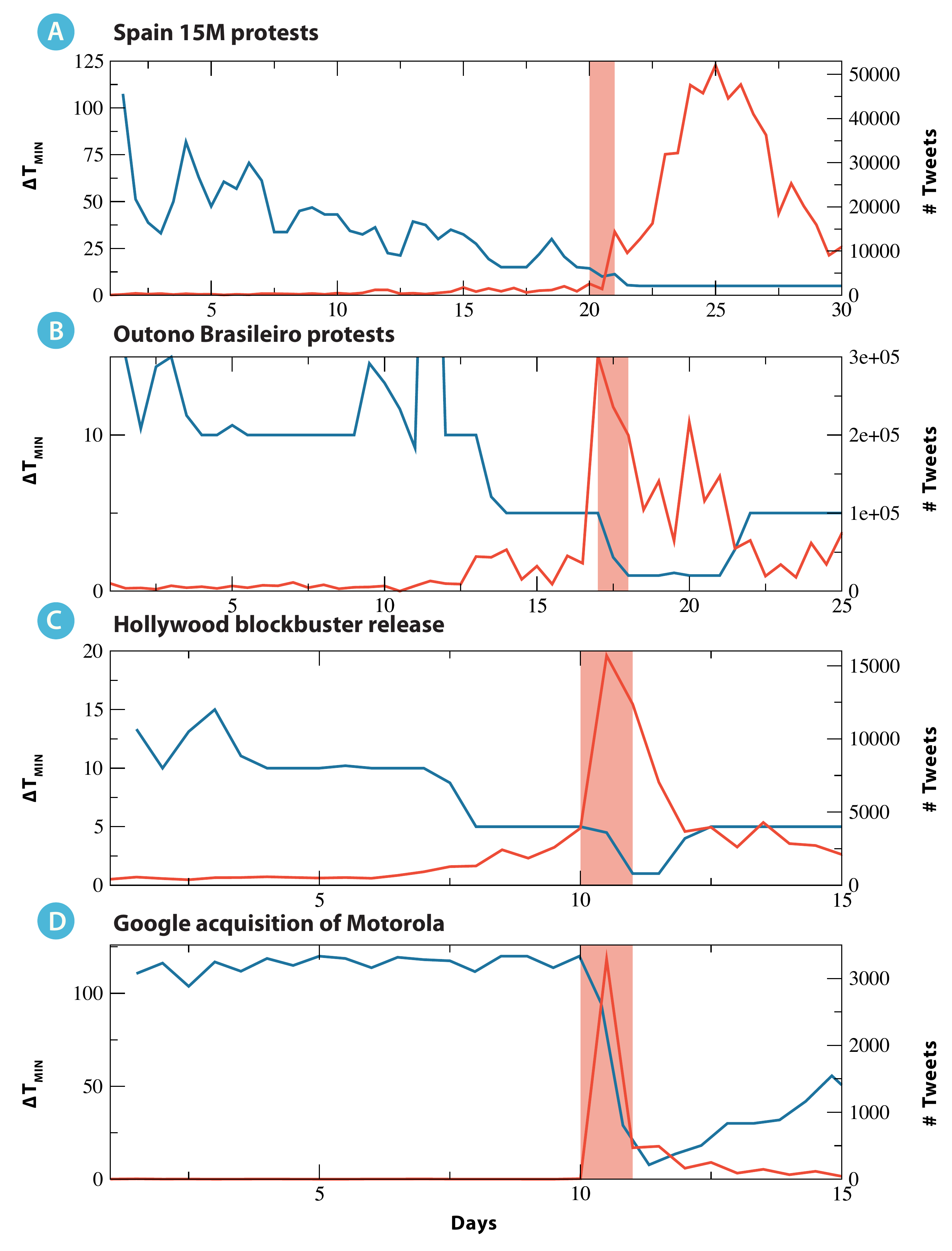}
\caption{In light blue we plot the characteristic time scale of STE for the A) 15M protests, B) Outono Brasileiro movements C) release of a Hollywood blockbuster D) the acquisition of Motorola by Google. We considered different geographical aggregation respect to figure 2 in the main text. The red lines show the activity in Twitter for each dataset, and the red strip indicate the day of the main collective event. }
\label{fig2_si}
\end{figure*}

\begin{figure*}[t]
\centering
\includegraphics[width=1.0\linewidth,clip=0]{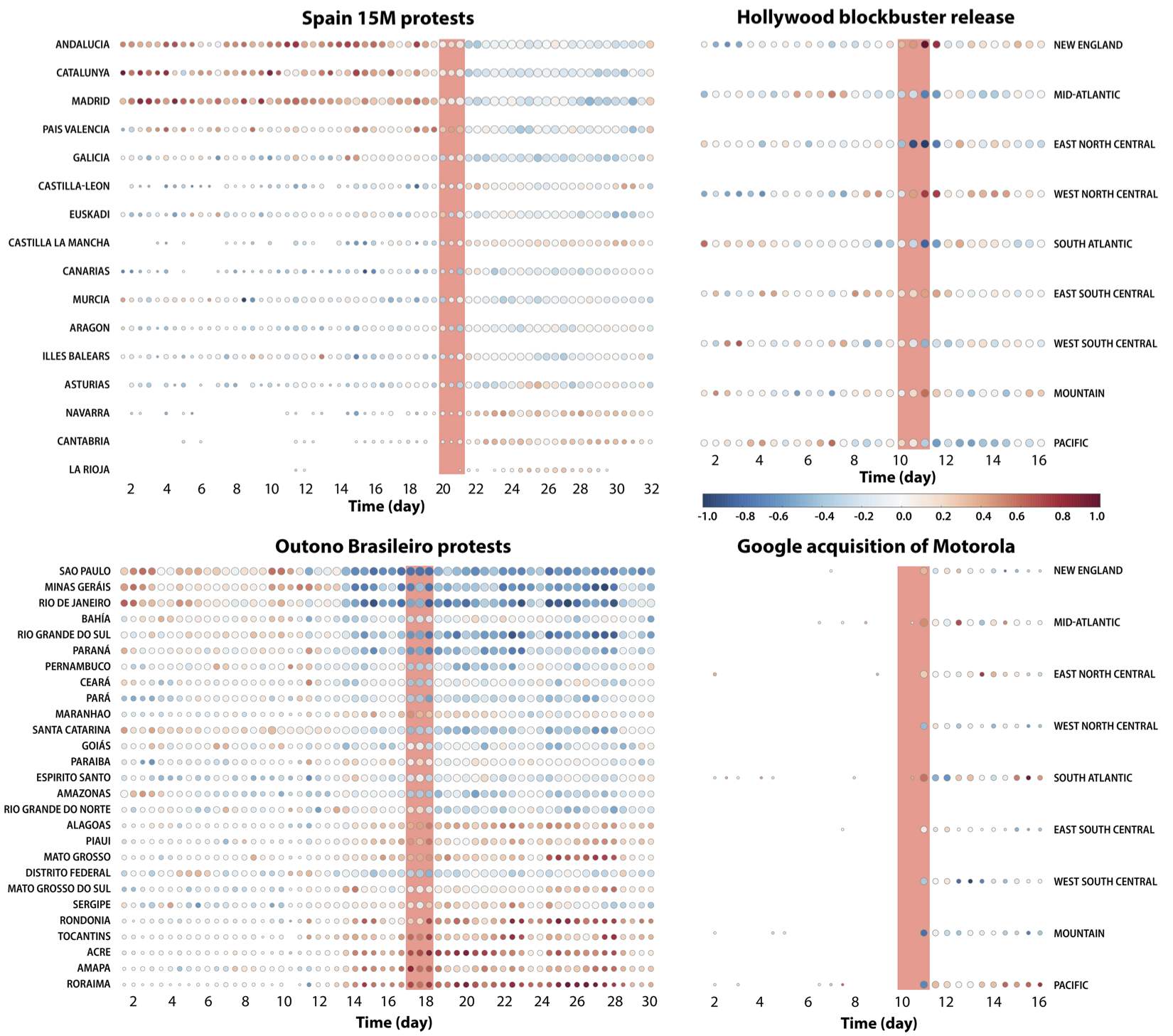}
\caption{We plot the normalized directionality index for the four datasets aggregated at different geographical levels respect to those used in the main text.}
\label{figure3_complete_different_geo}
\end{figure*}

\begin{figure*}[t]
\centering
\includegraphics[width=0.9\linewidth,clip=0]{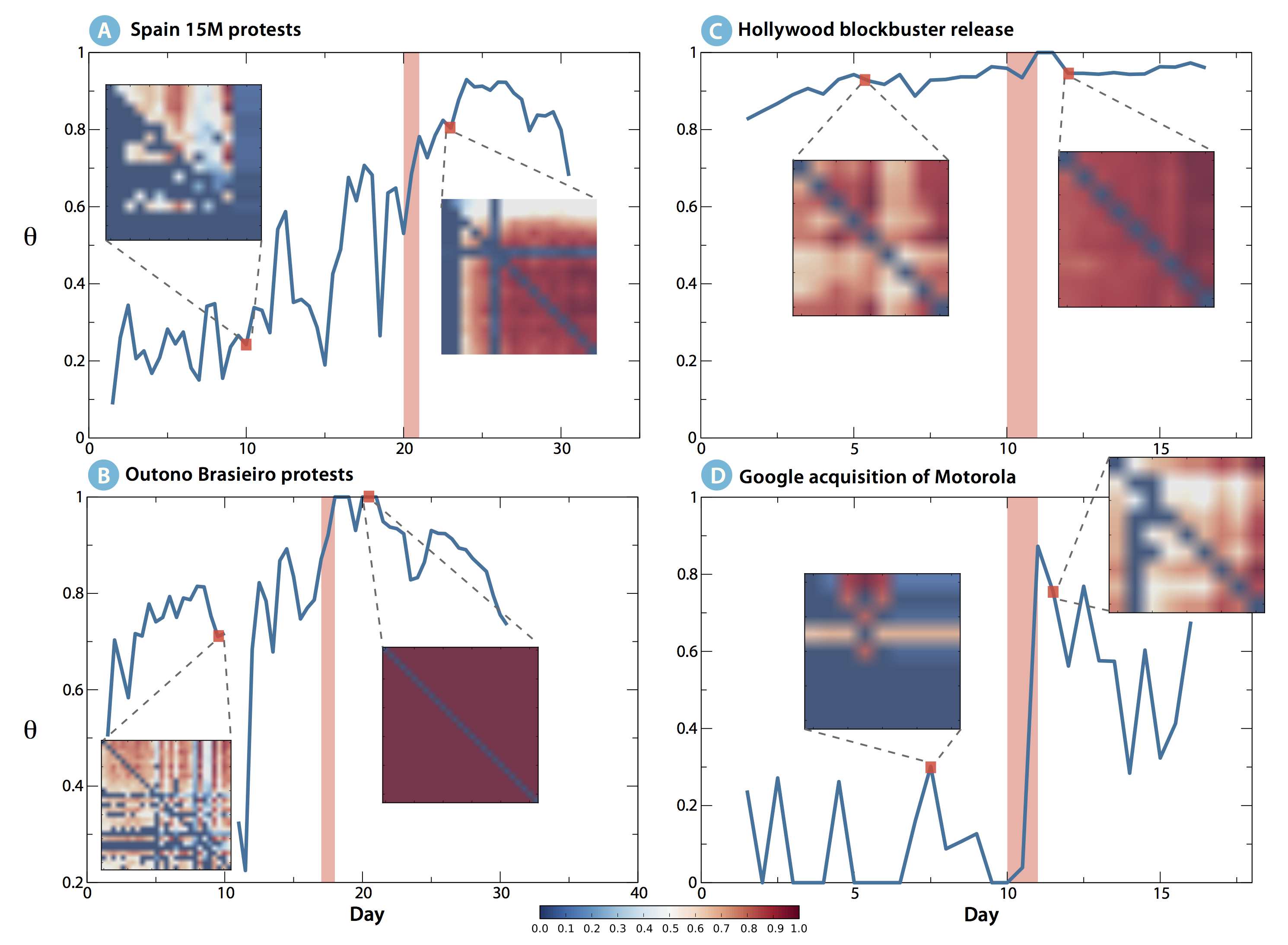}
\caption{Behavior of $\theta$ as a function of time for different geographical aggregations. For each dataset two matrices $T'$ are plotted considering a time before and one after the main event (signaled with a red vertical red bar).  A blurred transition can still be observed for events A and B. Note a point missing in the Brazilian dataset due to a data blackout between days 10 and 11.}
\label{space_new}
\end{figure*}

\section{Validation of results: controlled experiments}

In this section we validate our framework studying its performance on data surrogates, i.e. statistical ensembles of randomized  data. In other words, we apply our approach to a set of data that by construction do not contain the temporal correlations we find in real datasets. This step is crucial to prove that our observations capture genuine features of real collective events. In the following, for simplicity, we considered the 15M protests datasets.
\subsection{Statistical randomized surrogates of original data}
In order to validate our results, we need to make sure that our analysis and conclusions are mere artifacts which would arise in any case. To provide a reasonable baseline, we need to build randomized counterparts of the data and then analyze it just as we did for the actual case. In this line, we consider two methods to obtain data random surrogates: amplitude adjusted Fourier transform surrogates and constrained randomization surrogates, with an extensive use of the TISEAN software~\cite{tisean}.\\
We present the results for the randomization of the Spanish data, with qualitatively similar insights for the other datasets.

\subsubsection{Amplitude Adjusted Fourier Transform (AAFT) surrogates}
A first, robust step to provide a suitable null model is to generate randomized datasets which ensure that certain features of the original data will be preserved. In particular, we generate AAFT surrogates as proposed in \cite{schreiber1996improved}, who established an algorithm to provide surrogate datasets containing random numbers with a given sample power spectrum and a given distribution of values.

Under these constraints, we obtained 50 randomized versions of the 15M data, which were then analyzed in the same way as the original data (see main text and previous sections). The averaged results from such analysis are offered in Figure \ref{aaftsurr}. Clearly, the original patterns are completely blurred and just a single characteristic time scale can be observed. Furthermore, our approach do not capture any change in the the characteristic time scale as correlations and driving between different units have been artificially eliminated in the data.

\begin{figure}[h!]
\centering
\includegraphics[width=0.7\linewidth,clip=0]{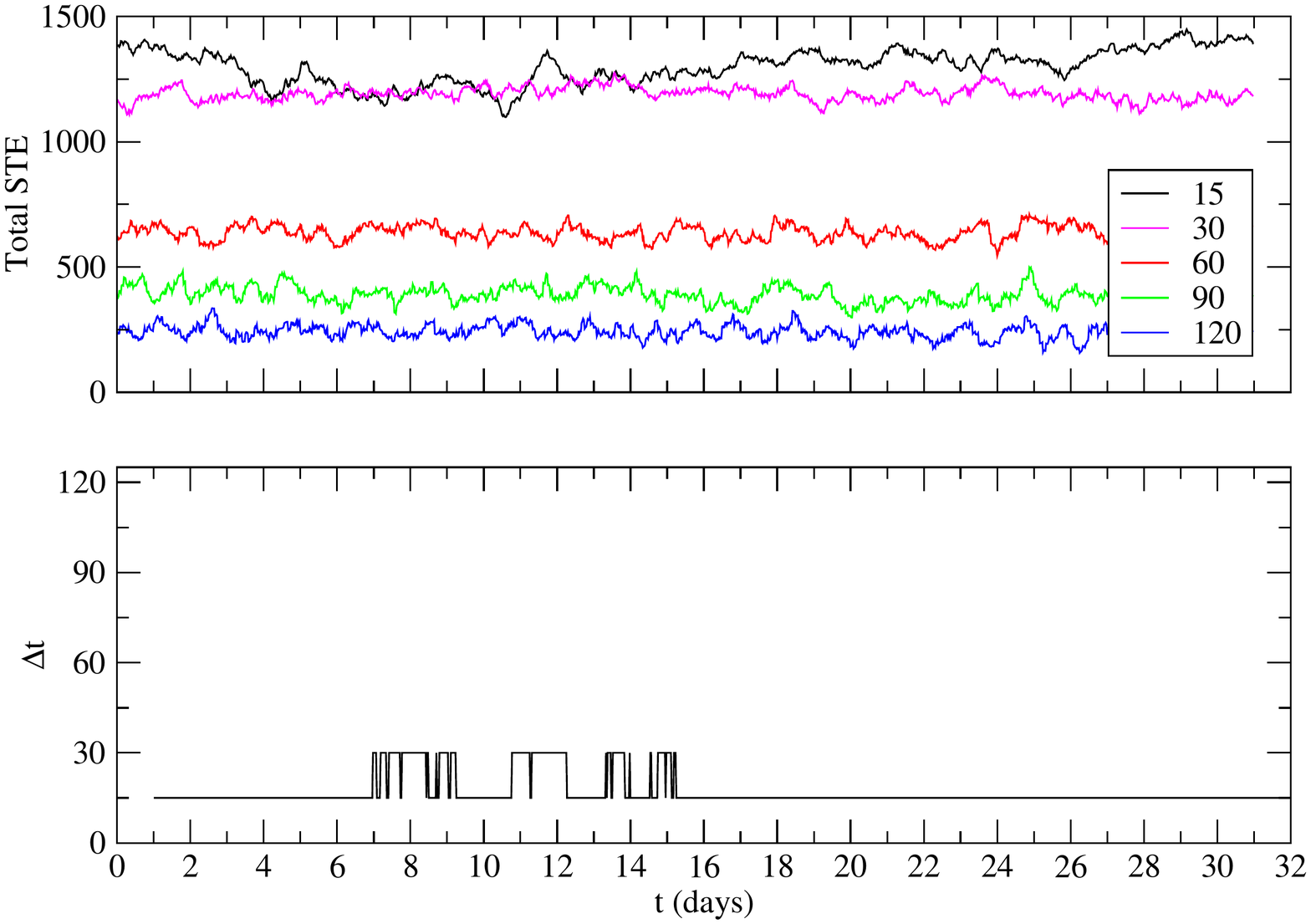}
\caption{Average total amount of STE for some $\Delta \tau$ (top panel) and time scale profile (bottom panel) for 15M AAFT surrogates (50 randomizations)}
\label{aaftsurr}
\end{figure}

Mirroring our analysis of real data, we intend to see whether some trace of the original transition is kept for this newly obtained random version of the data. To do so, we also exploit 50 randomizations of the original dataset, for which we can extract average surrogate snapshots (i.e. the state of the system at a given $t$ day). Just as in Figure 5 of the main text, Figure~\ref{fig4_SI_surro} shows two sorted (ranked) $T'$ matrices, corresponding to two different moments (before and after the main event) for the statistical randomized surrogates of original data. It is interesting to notice that as any localized abrupt change in the time-scale is washed out (Figure \ref{aaftsurr}), also any sort of systemic transition is missing.

\subsubsection{Constrained randomization surrogates}
Beyond a randomization scheme that guarantees given power spectrum and distribution of values, one might want to generate surrogates which are further constrained. This can be achieved if we demand randomized datasets to preserve as well a given non-periodic autocorrelation function (ACF). To this end, Schreiber \cite{schreiber1998constrained} developed a method of constrained randomization of time series data which seeks to meet the given constraints through minimization of a cost function, among all possible permutations, by the method of simulated annealing. 


\begin{figure}[h!]
\centering
\includegraphics[width=0.7\linewidth,clip=0]{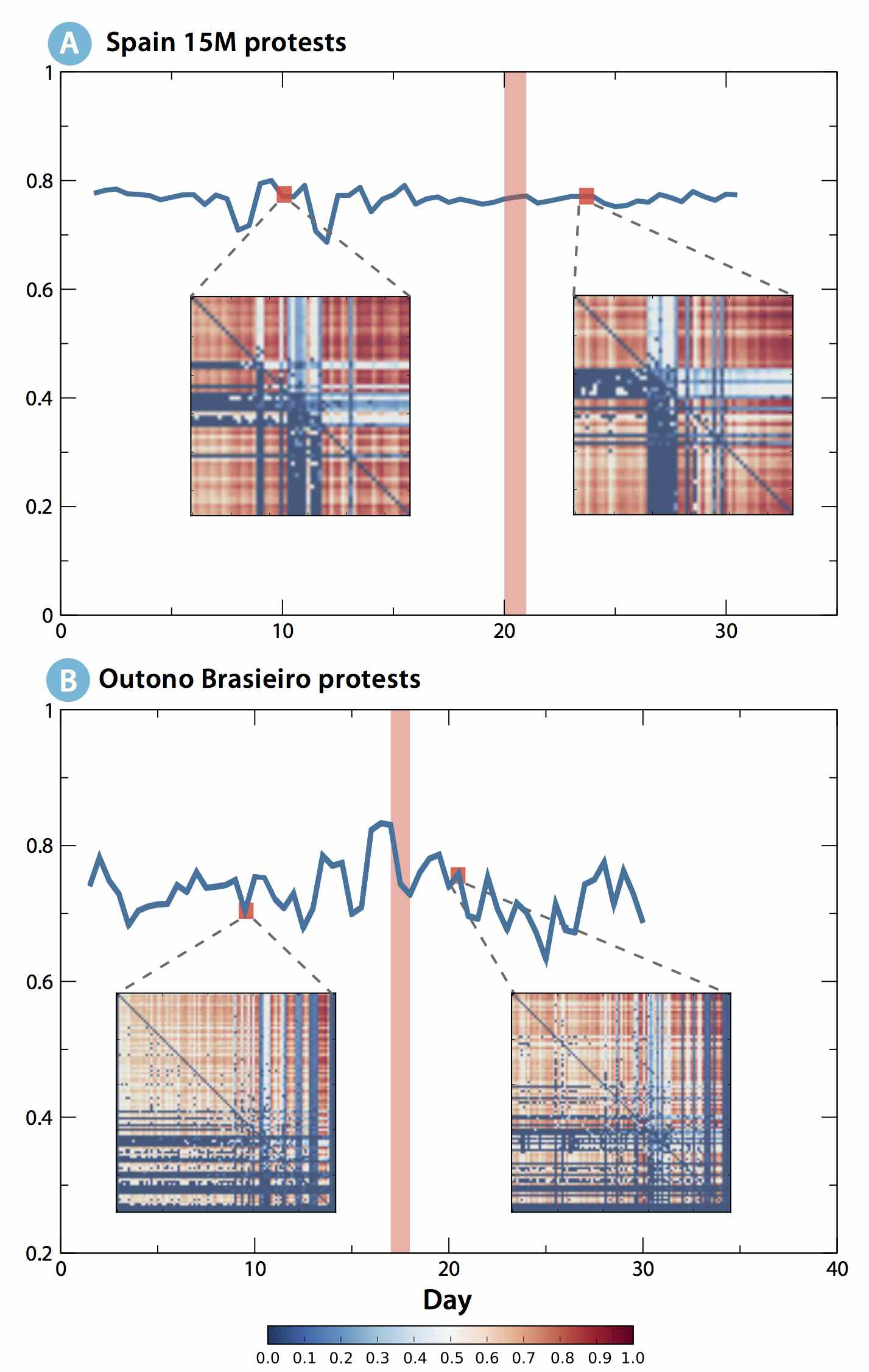}
\caption{Behavior of $\theta$ as a function of time for the statistical randomized surrogates of the 15M (A) and Brazilian dataset (B). For each dataset two matrices $T'$ are plotted considering a time before and one after the main event (signaled with a red vertical red bar). As expected in this context, no significant transition is observed in neither cases.}
\label{fig4_SI_surro}
\end{figure}

\begin{figure}[h!]
\centering
\includegraphics[width=0.7\linewidth,clip=0]{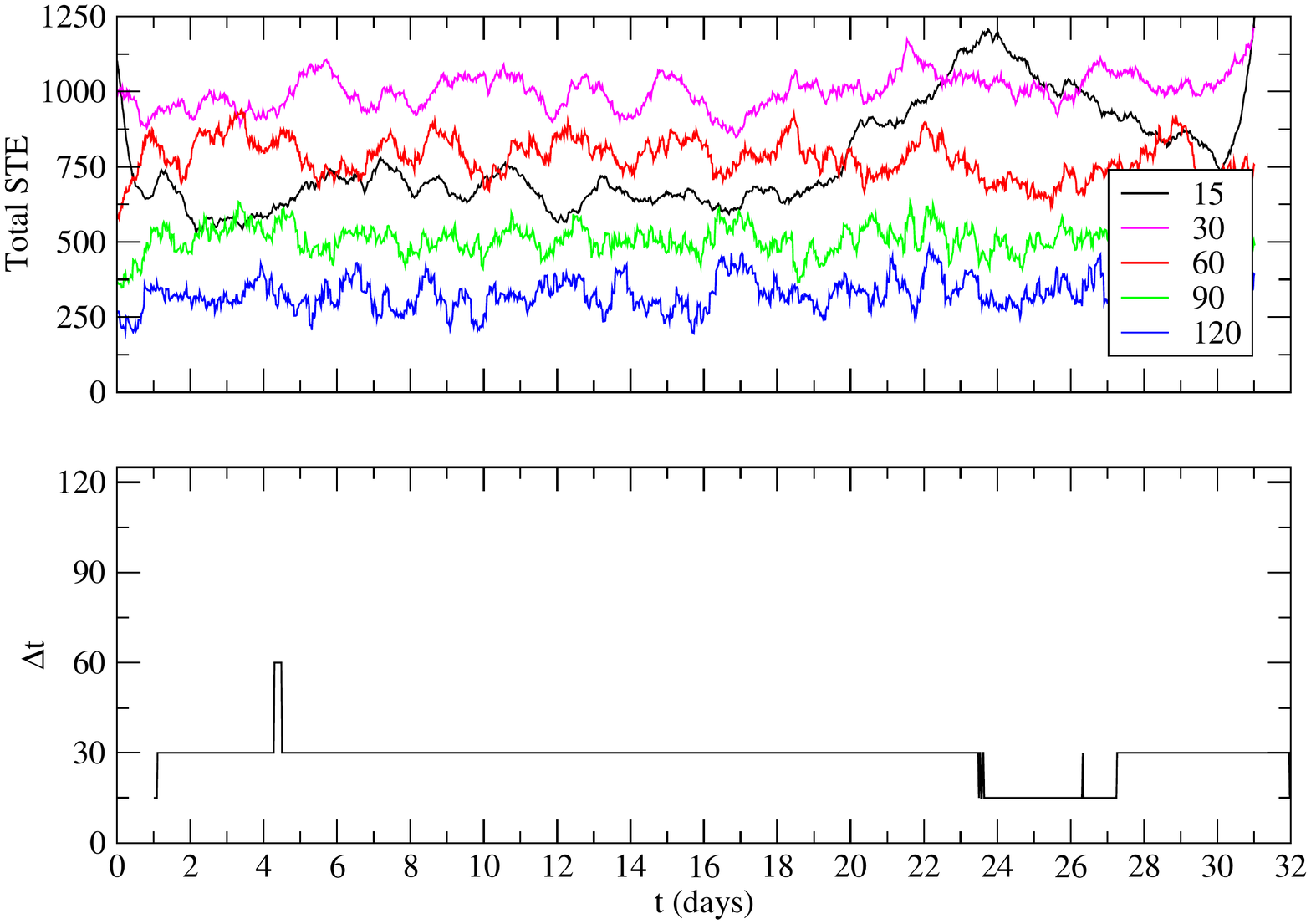}
\caption{Average total amount of STE for some $\Delta \tau$ (top panel) and time scale profile (bottom panel) for 15M constrained surrogates (20 randomizations)}
\label{constsurr}
\end{figure}

In Figure \ref{constsurr} the averaged results for the analysis of the constrained surrogate data can be checked. Twenty randomizations for the Spanish dataset were obtained. Even with the additional constraints (if compared with AAFT randomizations, see previous section), hardly any resemblance with the original patterns can be observed. It must be noted that constrained randomizations are time and CPU-consuming, due to the additional restrictions regarding ACF.

\begin{figure}[h!]
\centering
\includegraphics[width=\linewidth,clip=0]{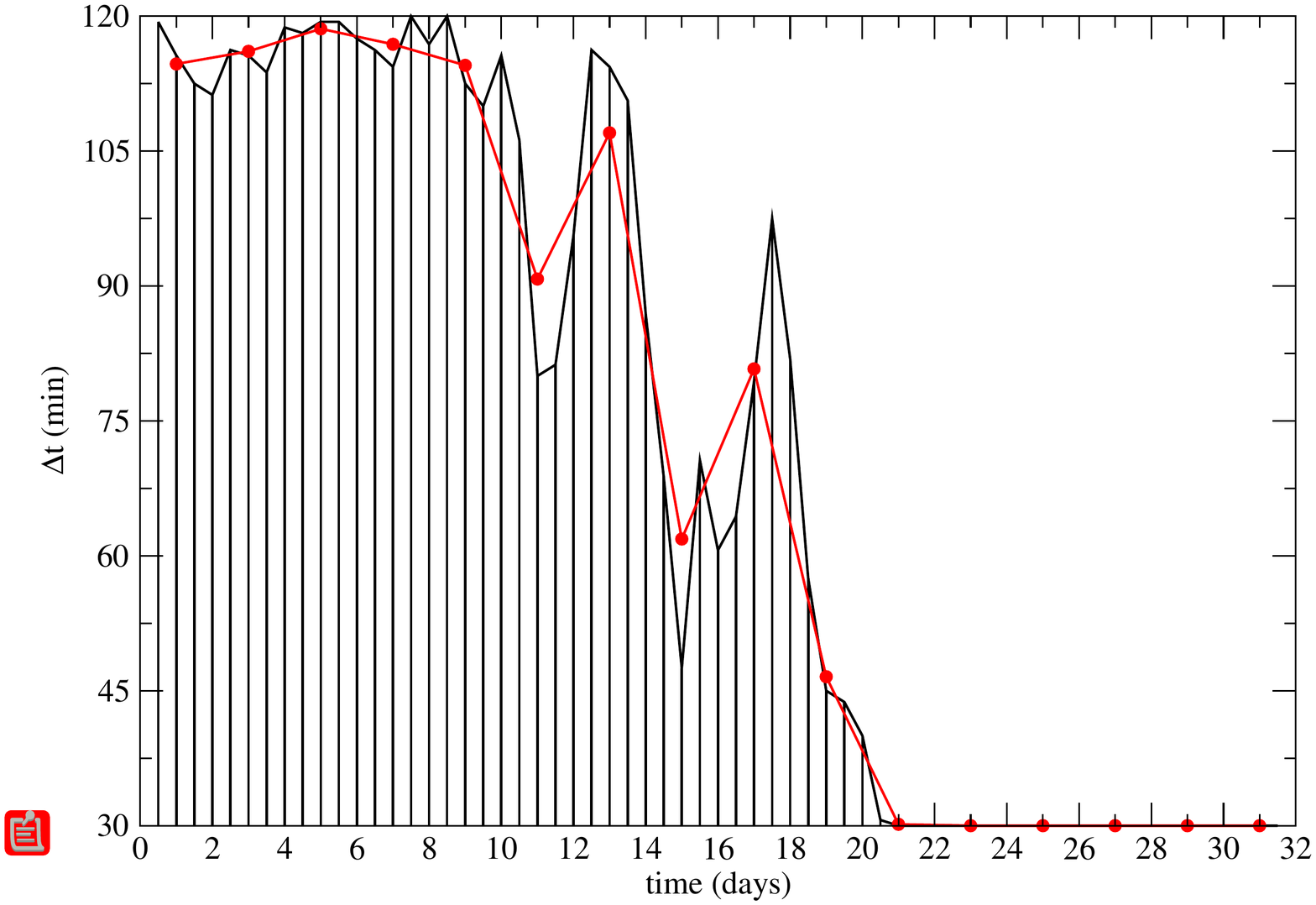}
\caption{Window width $w=1$}
\label{15Mts1d}
\end{figure}

\begin{figure}[h!]
\centering
\includegraphics[width=\linewidth,clip=0]{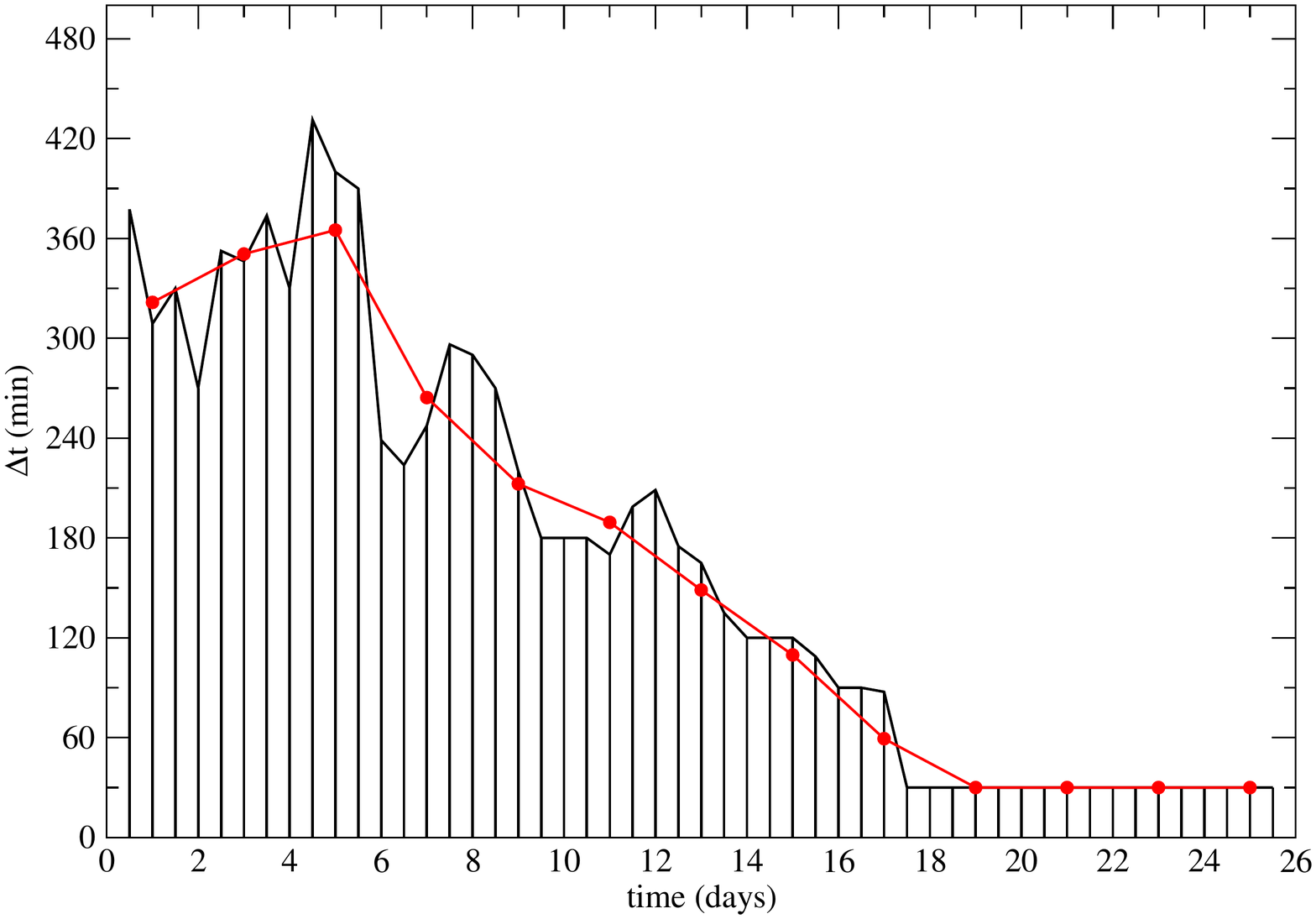}
\caption{Window width $w=7$}
\label{15Mts7d}
\end{figure}

\begin{figure}[h!]
\centering
\includegraphics[width=\linewidth,clip=0]{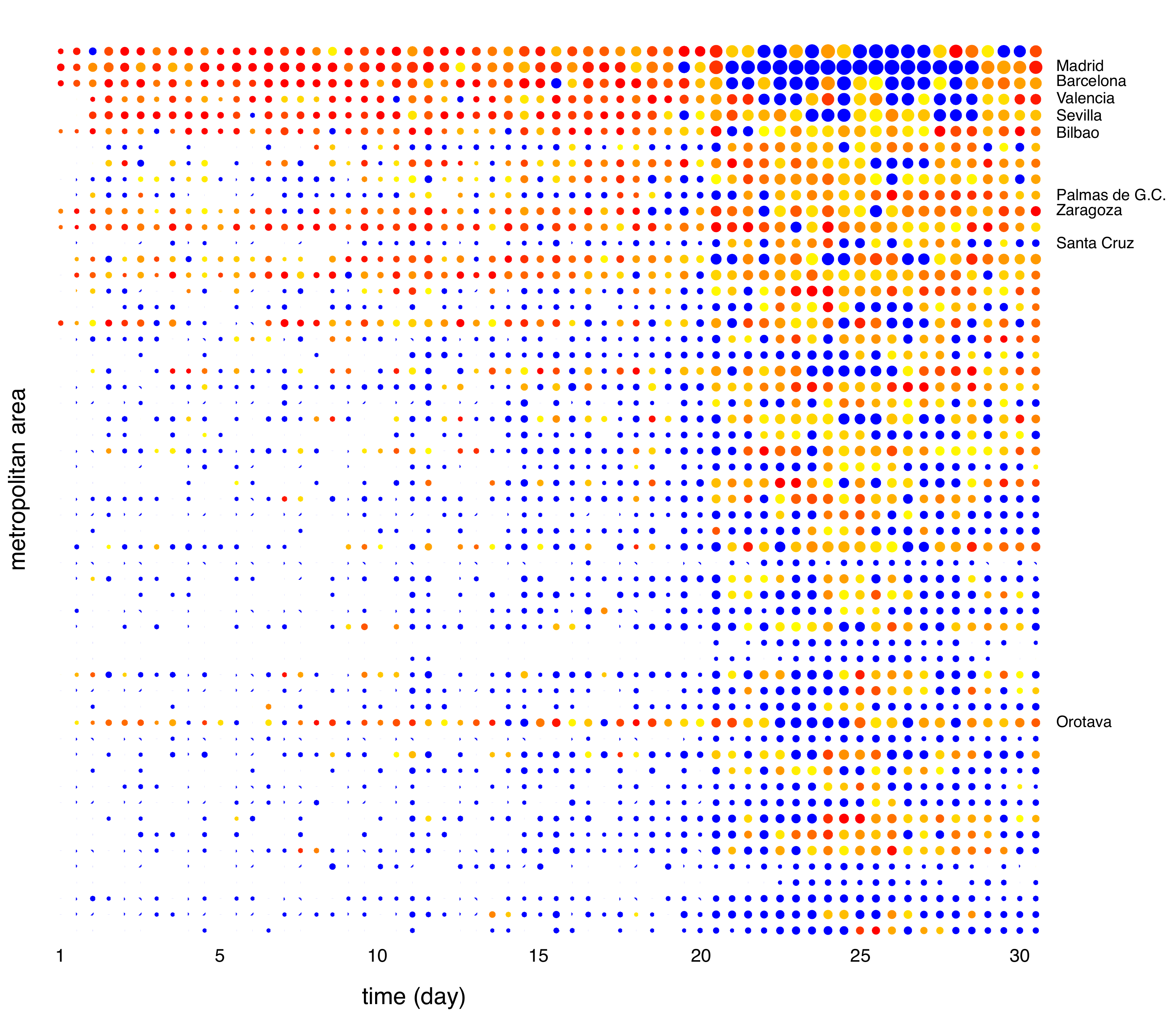}
\caption{Window width $w=1$}
\label{15Mb1d}
\end{figure}

\begin{figure}[h!]
\centering
\includegraphics[width=\linewidth,clip=0]{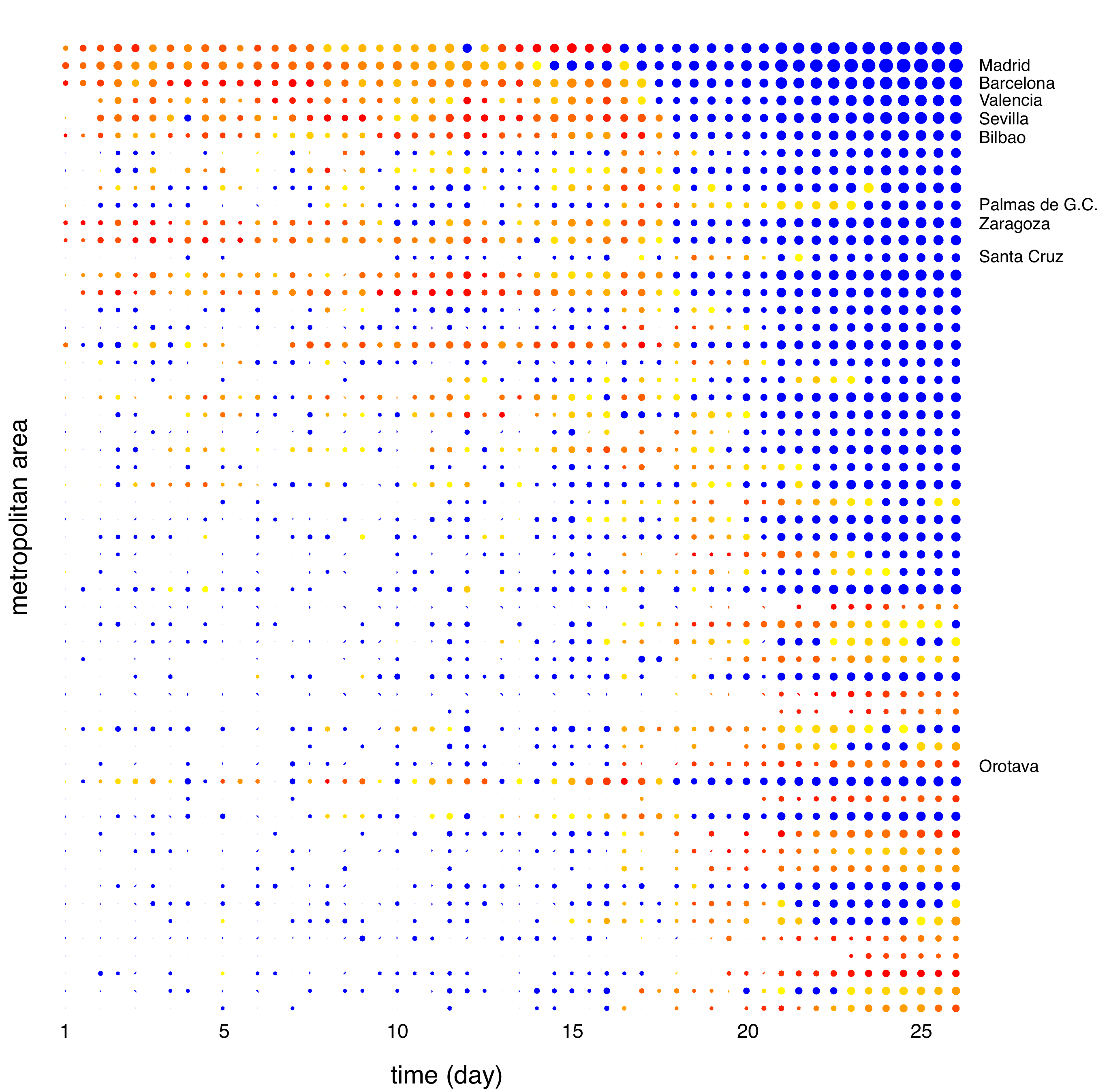}
\caption{Window width $w=7$}
\label{15Mb7d}
\end{figure}

\end{document}